\documentclass[lettersize,journal]{IEEEtran}
\usepackage{amsmath,amsfonts}
\usepackage{algorithmic}
\usepackage[ruled]{algorithm2e}  
\usepackage[caption=false,font=normalsize,labelfont=sf,textfont=sf]{subfig}
\usepackage{cite}
\usepackage{bm} 
\usepackage{makecell}
\usepackage{subcaption}
\usepackage{fmtcount}
\usepackage{array}
\usepackage[caption=false,font=normalsize,labelfont=sf,textfont=sf]{subfig}
\usepackage{textcomp}
\usepackage{stfloats}
\usepackage{url,lipsum}
\usepackage{verbatim}
\usepackage{graphicx}
\usepackage{mathrsfs}
\usepackage{booktabs}
\usepackage{multirow,url}
\usepackage[table]{xcolor}
\usepackage{bbding}
\usepackage{diagbox}
\usepackage{color}

\begin{document}

\title{Hardware-Friendly Plug-and-Play MVDR Beamforming for Microphone Arrays: Algorithm, Prototype, and System-Level Feasibility}

\author{Chengbo Chang,~\IEEEmembership{Student Member,~IEEE}, \, Ziye Yang,~\IEEEmembership{Student Member,~IEEE},\, Jie Chen,~\IEEEmembership{Senior Member,~IEEE}
\thanks{A preliminary version  has been accepted for International Conference on Acoustics, Speech and Signal Processing (ICASSP 2024).  This version provides a detailed algorithm description and designs a microphone beamforming system for further validation. }
\thanks{The authors are with Center of Intelligent Acoustics and Immersive Communications, Northwestern Polytechnical University, China(e-mail: ccb@mail.nwpu.edu.cn; yzy97@mail.nwpu.edu.cn; dr.jie.chen@ieee.org).}
\vspace{-5mm}
}

\markboth{Journal of \LaTeX\ Class Files,~Vol.~14, No.~8, August~2021}%
{Shell \MakeLowercase{\textit{et al.}}: A Sample Article Using IEEEtran.cls for IEEE Journals}


\maketitle

\begin{abstract}

Beamforming has been widely recognized for its spatial filtering capability in speech signal processing, enabling effective interference rejection and signal enhancement. The performance of adaptive beamforming methods, e.g., the conventional minimum variance distortionless response (MVDR) beamforming, often deteriorate significantly in complex and dynamic acoustic environments. Part of the reason is that such conventional beamforming ignores the intrinsic characteristics of speech. To address this issue, this paper proposes \textbf{\underline{P}}lug-a\textbf{\underline{n}}d-\textbf{\underline{P}}lay \textbf{\underline{M}}inimum \textbf{\underline{V}}ariance \textbf{\underline{D}}istortionless \textbf{\underline{R}}esponse (PnP-MVDR) algorithm, which integrates data-driven speech prior information into the MVDR framework to enhance robustness against non-stationary noise while maintaining its interpretability. The PnP-MVDR framework supports joint interaction of signal processing and deep learning methods and is designed to be compatible with modern DSP-NPU architectures, making it hardware-friendly for potential edge deployment. Beyond the algorithmic innovation, we have developed a complete microphone beamforming system to validate practical feasibility. Experimental results demonstrate that the proposed framework consistently outperforms traditional and deep learning-based baselines, establishing PnP-MVDR as an effective and deployable solution for edge audio scenarios.

\end{abstract}


\section{Introduction}
\IEEEPARstart{I}{n} real-world environments, target speech signals are frequently degraded by interfering sources, reverberation, and ambient noise, presenting formidable challenges for intelligent auditory systems such as mobile devices, smart speakers, and in-car voice assistants. These applications require speech processing algorithms that are optimized for embedded platforms to address high computational cost, limited robustness, and instability in complex environments~\cite{TCAS_SMI_MVDR_2008,TCAS_FPGA_SR_2015,TCAS_NeuromorphicPitch_2014}. While beamforming techniques effectively mitigate these issues, their performance often falters in complex acoustic settings, particularly on resource-constrained platforms that are unable to harness advanced algorithms for optimization~\cite{benesty2008microphone}. Consequently, there is a pressing need for robust microphone beamforming systems that not only enhance signal fidelity but also possess practical deployment capabilities~\cite{asr,zheng2019ultra,TCAS_ASR_2,TCAS_ASR_3,TCAS_AcousticSP_2020,TCAS_BinauralHA_2020,TCAS_DynamicFocusing_2020}. Extensive research has explored beamforming techniques, broadly categorized into fixed beamforming~\cite{das,fas,tong2022null,chen2021robustness,benesty2018fixed}, adaptive beamforming~\cite{mvdr,lcmv,TCAS_SMI_MVDR_2008}, and deep learning-based beamforming~\cite{fasnet,zhang2021adl,yoshioka2018multi,ochiai2020beam,stoller2018wave,lbt,qian2018deep}.

Fixed beamforming employs static filter weights, offering simplicity and reliability for hardware implementation due to its minimal computational complexity and low circuit overhead~\cite{benesty2018fixed,TCAS_AcousticSP_2020,TCAS_BinauralHA_2020}. Delay-and-sum beamforming (DSB) enhances target signals by compensating for time differences of arrival (TDOA) across microphone array elements, achieving coherent superposition but with limited spatial resolution~\cite{das}. Maximum directivity factor (MDF) refines array weights to focus the beam pattern, enhancing noise suppression~\cite{benesty2018fixed}. Superdirectional beamforming (SDB) leverages an isotropic noise field to narrow the main lobe at low frequencies~\cite{chen2021robustness}. Null steering beamforming creates nulls in interference directions for enhanced suppression but lacks adaptability to dynamic environments~\cite{tong2022null}. Despite their hardware efficiency, fixed methods struggle in varying acoustic conditions, necessitating adaptive approaches that balance algorithmic flexibility with circuit constraints, as demonstrated in neuromorphic noise reduction~\cite{TCAS_NeuromorphicPitch_2014} and flexible microphone array designs~\cite{fikes2021framework}.

Adaptive beamforming dynamically adjusts filter coefficients based on signal characteristics, delivering superior performance and physical interpretability ideal for hardware acceleration~\cite{chen2007adaptive,chen2008adaptive}. MVDR beamformer minimizes output signal energy while preserving gain in the target direction, effectively suppressing interference~\cite{mvdr,TCAS_SMI_MVDR_2008}. The linearly constrained minimum variance (LCMV) beamformer extends MVDR with multiple constraints, enabling precise control in complex scenarios~\cite{lcmv}. Both methods face challenges related to robustness against array imperfections and dynamic signal environments, often necessitating techniques like diagonal loading or regularization to maintain performance. In addition, these methods often neglect intrinsic speech characteristics, limiting their efficacy and prompting hybrid integrations with data-driven priors, as evidenced in low-latency phase-sensitive masking~\cite{TCAS_BinauralHA_2020} and wideband direction-of-arrival (DOA) estimation~\cite{lin2023wideband}.


Deep learning has transformed speech signal processing through the nonlinear modeling capabilities of deep neural networks (DNNs). Neural beamforming integrates DNNs into the beamforming process, categorized into filtering-based (FB), masking-based (MB), and regression-based (RB) approaches~\cite{fasnet}. FB methods learn beamforming filters in time or frequency domains~\cite{fasnet,zhang2021adl}, while MB methods estimate time-frequency masks to compute spatial covariance for beamformers like MVDR~\cite{yoshioka2018multi,ochiai2020beam,lbt}. RB methods implicitly embed beamforming within DNNs, mapping multi-channel inputs to target sources~\cite{stoller2018wave}. Hybrid strategies, such as single-channel enhancement with wiener filtering, further enhance performance~\cite{qian2018deep}. Yet, these approaches often operate as black-box systems, lacking interpretability and posing challenges for hardware optimization due to high power and area demands in multi-layer DNNs~\cite{abdullah2024hardware,zheng2019ultra,liu2021more,yang202471,xie2021dyga}.

To overcome these challenges, we propose the Plug-and-Play MVDR (PnP-MVDR) algorithm, which integrates data-driven speech priors into the MVDR framework via the regularization-by-denoising (RED) strategy~\cite{chen2023integration,wangxiuheng,yang2023integrating,pnp_iva,chang2024plug}. This approach reformulates MVDR optimization with a DNN-based denoiser as a regularizer, ensuring mathematical rigor and alignment with speech signal characteristics while leveraging scalable MVDR techniques for low-power covariance estimation and embedded processing~\cite{TCAS_SMI_MVDR_2008,TCAS_NeuromorphicPitch_2014,abdullah2024hardware,yang202471}.
The proposed algorithm effectively integrates the advantages of beamforming technology with a DNN-based denoiser, while supporting a DSP-NPU architecture wherein the DSP manages the signal processing pipeline and the NPU accelerates DNN-based prior extraction~\cite{TCAS_AcousticSP_2020,TCAS_NLU_2025}. Thus, to ensure hardware compatibility and practical deployment, we construct a complete microphone beamforming system, which demonstrates superior signal quality compared to traditional and neural baselines in both simulated and real-world scenarios~\cite{kiselev2022spiking,bos2023end,alsakkal2025spiketrum}.

The main contributions of this paper are summarized as followings:
\begin{itemize}
\item  PnP-MVDR algorithm: We propose a novel Plug-and-Play MVDR framework that incorporates data-driven speech priors to enhance robustness against non-stationary noise while preserving the interpretability of conventional beamforming.
\item Joint signal processing and deep learning integration: The framework supports a flexible interaction between classical beamforming and neural network-based priors, designed to be compatible with modern DSP-NPU architectures, enabling hardware-friendly edge deployment potential.
\item Hardware system validation: A complete microphone beamforming system is developed to demonstrate practical feasibility, showing that the proposed method can work effectively with real captured signals.
\end{itemize}

The remainder of this paper is organized as follows. 
In Section II, we describe the speech signal model and review several widely-used beamforming techniques.
In Section III, our proposed PnP-MVDR algorithm is thoroughly derived. The DSP-NPU-supported microphone beamforming system based on this algorithm is described in detail in Section IV.
Section V provides the experiment setup including simulated data and real-world data.
We present and analyze experimental results in Section VI. 
Section VII concludes this paper.

\section{Signal Model and Beamforming Methods}

\subsection{Signal Model}
We assume the signal model under a scenario which contains a target speaker, an inference speaker and a microphone array with $Q$ channels. In the time domain, the mixed speech signal received by $q$-th channel is expressed by:
\begin{equation}
\label{sig_t}
x_q(t)=h_{1q}(t)*s_1(t)+h_{2q}(t)*s_2(t)+z_q(t),
\end{equation}
with $s_1(t)$ and $s_2(t)$ representing the source speech of two different speakers. $h_{1q}(t)$ and $h_{2q}(t)$ are the time-invariant acoustic transfer functions~(ATFs) from two speech sources to $q$-th channel, respectively. In addition, we consider the zero-mean additive noise that is independent of the speech sources, denoted by $z_q(t)$ in (\ref{sig_t}). Approximating signal model (\ref{sig_t}) in short-time Fourier transform (STFT) yields:
\begin{equation}
\label{sig_tf}
\begin{split}
    X_q(n,k) =& \sum_{j=0}^{J-1}H_{1q}(j,k)S_1(n-j,k) \\  \\[-25pt]
    &\qquad + \sum_{j=0}^{J-1}H_{2q}(j,k)S_2(n-j,k) +Z_q(n,k),
\end{split}
\end{equation}
where $n$ and $k$ denote the time-frame and frequency bin indices
respectively. With $J$ being the order of ATFs, $H_{iq}(j,k)$ and $S_i(n,k)$ represent the counterparts of $h_{iq}(t)$ and $s_i(t)$ of speaker $i$ (with $i\in\{1, 2\}$) in STFT domain respectively.

Regarding Speaker $1$ as the target speaker, the task comes to estimate $\bm{S}_1$\footnote{Note that $\bm{S}_1\in \mathbb{C}^{N\times K}$ is defined as the speech time-frequency matrix, and the similar definition will be used for the other bold capital letters.} from the observed speech of microphone array, i.e., ${\bm X}$ $\in \mathbb{C}^{Q\times N\times K}$ stacked by $\{\bm{X}_q\}_{q=1}^{Q}$. Generally, Our objective is to design a filter for estimating the target speaker's signal:
\vspace{1mm}
\begin{equation}
\label{filter}
\widehat{\bm{S}}_i=\mathcal{F}({\bm X}),
\end{equation}
where $ \mathcal{F}$ is the mapping function represented by the designed filter. Similarly, the same processing can be applied to the other speaker to achieve the goal of separation. From the perspective of beamforming, the problem reduces to computing an optimal set of weighting coefficients.

\subsection{Conventional Adaptive Beamforming}
Beamforming utilizes spatial selectivity to suppress background noise and interference from other speakers to obtain an estimated target signal. Considering that the speech signal is a broadband signal, the  beamformer is required to obtain frequency-dependent complex-valued weight vectors $ \mathbf{w}(k)=[w_1(k),\ldots,w_Q(k)]^T $. Following the form of (\ref{filter}), the output of the beamformer for the $k$-th band is given by
\begin{equation}
\label{beam}
\widehat{\bm{S}}_i(k)=\mathcal{F}({\bm X(k)};\mathbf{w}(k))=\mathbf{w}(k)^\mathrm{H}{\bm X}(k),
\end{equation}
where $(\cdot)^{\rm H} $ represents the conjugate transpose, $ {\bm{X}}(k)\in\mathbb{C}^{Q\times N}  $ is the spectrum corresponding to the $k$-th band of observations. Compared to the fixed beamformer, adaptive beamformer is more robust as its weight vectors are dynamically adapted to the signal, and one such method is MVDR beamformer.

The MVDR beamformer aims to minimize the energy of the output signal while maintaining the gain in the target direction, effectively suppressing interference signals. This optimization problem is mathematically formulated as:
\begin{equation}
\label{mvdr}
\begin{aligned}
    &\min\enspace\frac{1}{N}\sum_{k=1}^K\| \mathbf{w}^{\rm H}(k){\bm{X}}(k)\|_2^2\\
    &\text{s.t.}\quad \mathbf{w}^{\rm H}(k)\bm{{a}}(k)=1,\quad\enspace \text{for}~k=1,...,K,
\end{aligned}
\end{equation}
where $\bm{a}(k)$ denotes the direction vector of the $k$-th band for target speech, which is crucial to ensure that the desired signal is not distorted. For a signal with an azimuth of $\theta$, its direction vector can be represented as:
\begin{equation}
\label{a(k)}
\bm{a}\left(k, \theta\right)=\begin{bmatrix}1 & e^{-j2\pi k\frac{\delta cos\theta}{c}} \ldots & e^{-j2\pi k\frac{(Q-1)\delta cos\theta}{c}}\end{bmatrix}^{T}, 
\end{equation}
where $\delta$ and $c$ denotes the inner distance of microphones and the speed of sound respectively. Note that if not known, $\theta$ can be an estimate of DOA of the target source.

We directly write the Lagrangian form of problem (\ref{mvdr}) as follows:
\begin{equation}
\label{L_mvdr}
\begin{aligned}
    \mathcal{L}( {\mathbf{w}}(k),{\bm{\lambda}}(k)) = &\frac{1}{N}\sum\limits_{k=1}^K \| {\mathbf{w}^{\rm H}}(k){\bm{X}}(k)\| _2 ^2\\
    & + \sum\limits_{k=1}^K{\bm \lambda}(k)\left(\mathbf{w}^{\rm H}(k)\bm{a}(k)-1\right),
\end{aligned}
\end{equation}
where  $\bm \lambda(k)$ denotes the Lagrange multiplier. By setting the partial derivatives of $\mathbf{w}(k)$ and $\bm \lambda(k)$ to zero respectively, we can obtain the optimal weight vector of the MVDR beamformer:
\begin{equation}
\label{w_mvdr}
\mathbf{w}_{\text{opt}}\left(k\right)=\frac{\mathbf{R}_{\bm{x}}^{-1}\left(k\right)\bm{a}(k)}{\bm{a}^{\rm H}\left(k\right)\mathbf{R}_{\bm{x}}^{-1}\left(k\right)\bm{a}(k)},
\end{equation}
where $\mathbf{R}_x(k)=\frac{1}{N}{\bm{X}}(k){\bm{X}}^{\rm H}(k)$ represents the covariance matrix of observations. Following equation (\ref{beam}), we can obtain an estimate for each frequency of the target speech:
\begin{equation}
\label{s_mvdr}
\widehat{\bm{S}}_i(k)=\mathbf{w}_{\text{opt}}(k)^\mathrm{H}{\bm X}(k).
\end{equation}

Explicitly, the MVDR beamformer has a clear physical interpretability and can work steadily in complex environment. In addition, it can be shown from the analytical solution that the direction vector and the covariance matrix directly affect the speech separation effect, the former of which we assume to be known or obtained by DOA estimation, so the latter becomes a key point in optimizing the MVDR beamformer. 


\subsection{Speech-adaptive Beamforming}
Although the MVDR method offers a clear physical interpretation, its formulation does not explicitly account for the fact that the beamformer output should exploit the intrinsic properties and statistical priors of speech signals. Consequently, this universal physical model falls short in capturing the unique characteristics of speech when applied as a spatial filtering technique. Speech-adaptive beamforming addresses this limitation by incorporating speech priors derived from the probabilistic characteristics of speech distributions.

It is well established that speech coefficients in the short-time Fourier transform (STFT) domain, when modeled as random variables, can be approximated by a zero-mean circularly symmetric complex gaussian (CG) distribution. The maximum likelihood distortionless response (MLDR) beamformer~\cite{gaussian} is specifically designed so that its output follows such a CG prior distribution, with time- and frequency-dependent variances that reflect the inherently non-stationary nature of speech signals. Accordingly, the probability density function (PDF) of the MLDR beamformer output is expressed as
\begin{equation}
\mathcal{P}(\widehat{\bm{S}}_i(n,k))=\frac{1}{\pi\gamma(n,k)}e^{-\frac{|\widehat{\bm{S}}_i(n,k)|^2}{\gamma(n,k)}},
\label{pdf}
\end{equation}
where $\widehat{\bm{S}}_i(n,k)$ and $\gamma(n,k)$ represents the output of beamformer and the variance of the prior distribution at the $n$-th frame and $k$-th band respectively. By applying the principle of logarithmic maximum likelihood, the corresponding Lagrangian function can be formulated as follows:
\begin{equation}
\label{eq:lag_gau}
\begin{aligned}
\mathcal{L}(&\mathbf w(k),\gamma(n,k))=\sum_{n=1}^{N} \frac{\mathbf{w}^{\rm H}(k)\mathbf{X}(n,k)\mathbf{X}^{\rm H}(n,k)\mathbf{w}(k)}{\gamma(n,k)} + \\
&\sum_{n=1}^{N}\left(\log(\pi)+\log \gamma(n,k)\right)+\alpha_k(\mathbf{w}^\mathrm{H}(k)\mathbf{a}(k,\theta)-1),
\end{aligned}
\end{equation}
where $\widehat{\bm{S}}_i(n,k) = \mathbf{w}^{\rm H}(k)\mathbf{X}(n,k)$ has been substituted, $\alpha_k$ is the Lagrangian multiplier. The estimate of the weight vector can then be obtained by iteratively differentiating the cost function.

Motivated by the sparse nature of speech coefficients in the STFT domain, \cite{laplacian} further modeled the real and imaginary parts of the beamformer output as random variables following zero-mean Laplacian distributions:
\begin{equation}
\mathcal{P}(\widehat{\bm{S}}_i(n,k))=\frac{1}{\gamma(n,k)}e^{-2\frac{|\mathcal{R}(\widehat{\bm{S}}_i(n,k))|+|\mathcal{I}(\widehat{\bm{S}}_i(n,k))|}{\sqrt{\gamma(n,k)}}},
\end{equation}
where $ \mathcal{R}(\cdot)$ and $\mathcal{I}(\cdot)$ denote the real and imaginary parts of a complex number respectively. In analogy to equation~(\ref{eq:lag_gau}), the corresponding Lagrangian function can be formulated and solved using the CVX optimization toolkit~\cite{toolkit}.

\section{The Proposed Algorithm}
While the aforementioned speech-adaptive beamforming techniques incorporate speech priors by modeling the PDF of speech signals, the intrinsic complexity of speech characteristics often surpasses what can be effectively captured through distributional modeling alone. To address this limitation, our goal is to design a beamformer specifically tailored for speech signals that can leverage deep speech priors, thereby substantially enhancing the performance of the conventional MVDR method.
In this section, we first reformulate the MVDR beamforming optimization problem to integrate deep speech priors, and then introduce the proposed solution approach.

\subsection{Problem Reformulation}
Considering that beamforming based spatial selectivity, it is vulnerable to the influence of uncorrelated noise and interference signals in the target direction. Therefore, we model the desired signal speech signal of each frequency as:
\begin{equation}
\label{disire}
{\bm R(k)} = \mathbf{w}^{\rm H}(k){\bm{X}}(k)-\bm{V}(k).
\end{equation}
where  $\bm{V}(k)$ is an estimate of noise and interference during modeling and processing. The rationale for subtracting this term is discussed in~\cite{yang2023integrating}, and its validity is further verified in our experimental results. To design a customized beamformer, the filter must be tailored to the problem formulation such that the desired signal is explicitly treated as speech rather than as an arbitrary source. To this end, we incorporate speech priors by introducing a regularization term into the optimization problem formulated in~(\ref{mvdr}), thereby constraining the output to exhibit the intrinsic structural characteristics of speech signals. Since ${\bm R} \in \mathbb{C}^{N \times K}$, formed by stacking ${\bm{R}(k)}_{k=1}^{K}$, represents the estimated desired speech signal, introducing such regularization is beneficial for embedding speech priors, leading to the following formulation:
\begin{equation}
\label{pnp1}
\begin{aligned}
    &\min\enspace\frac{1}{N}\sum_{k=1}^K\|  \mathbf{w}^{\rm H}(k){\bm{X}}(k)\|_2^2+{\alpha}\mathcal{J}_{\text{Reg}}({\bm{R}}),\\
    &\text{s.t.}\quad \mathbf{w}^{\rm H}(k)\bm{a}(k)=1,\quad\enspace \text{for}~k=1,...,K,
\end{aligned}
\end{equation}
where $\alpha$ represents the trade-off parameter, and $\mathcal{J}_{\text{Reg}}$ denotes a regularizer. 

However, handcrafting an effective regularizer $\mathcal{J}_{\text{Reg}}$ and developing its optimization solver presents significant challenges. To address this, we propose to learn priors from speech data and utilize the PnP strategy to incorporate the deep priors into the mathematics-based optimization. Specifically, we consider $\mathcal{J}_{\text{Reg}}$ in the form of RED \cite{red}, a principled methodology for transforming denoisers into explicit regularizers that effectively capture deep priors. In addition, RED also has favorable derivative properties under mild assumptions. The specific RED-based regularization form in our framework takes the form: 
\begin{equation}
\label{reg}
\mathcal{J}_{\text{Reg}}({\bm R}) = \frac{1}{2}\langle \,\bm{R}, \bm{R} - \Omega(\bm{R}) \,\rangle
\end{equation}
where $\Omega(\cdot)$ denotes an off-the-shelf denoiser. The output of the denoiser is naturally characterized by speech, so the present method can treat the output as a speech signal by adding such a penalty term to the original problem. Notably, the penalty itself is proportional to the inner product between the desired speech $ R$ , and its denoised residual $ {\bm R}-\Omega({{\bm R}}) $ . This interpretation aligns with a speech-adaptive Laplacian.

\subsection{Solving Method}
After introducing a regularization term with respect to $ \bm{R} $, the complete form of the optimization problem can be written as:
\begin{equation}
\label{pnp2}
\begin{aligned}
    \min\limits_{{\mathbf{w}}(k),\bm{R},\bm{V}}\enspace&\frac{1}{N}\sum\limits_{k=1}^K\left\|  {\mathbf{w}}^{\rm H}(k){\bm{X}}(k)\right\| _2 ^2+\frac{\alpha}{2}\langle \,\bm{R}, \bm{R} - \Omega(\bm{R}) \,\rangle\\
    \text{s.t.}\quad &\bm{R}(k)={\mathbf{w}}^{\rm H}(k){\bm{X}}(k)-\bm{V}(k)\\
    &\mathbf{w}^{\rm H}(k)\bm{a}(k)=1,\quad\enspace \text{for}~k=1,...,K
\end{aligned}
\end{equation}

To solve this problem, we alternately use Lagrange and augmented Lagrange techniques on the two constraints. Specifically, we primarily employ augmented Lagrange technique for the first constraint because it brings robustness and yields convergence without assumptions like strict convexity or finiteness~\cite{boyd2011distributed}. Then we obtain the (scaled) augmented Lagrange function of the problem:
\begin{equation}
\label{L1}
\small
\begin{aligned}
    \mathcal{L}( {\mathbf{w}}(k),&\bm{R},\bm{V},\bm{P}) =\frac{1}{N}\sum\limits_{k=1}^K\left|\left|  \mathbf{w}^{\rm H}(k){\bm{X}}(k)\right|\right| _2 ^2+\frac{\alpha}{2}\langle \,\bm{R}, \bm{R} - \Omega(\bm{R}) \,\rangle\\
    &+\frac{\rho}{2}\sum\limits_{k=1}^K\left(\left\|\mathbf{w}^{\rm H}(k){\bm{X}}(k)-\bm{R}(k)-\bm{V}(k)+\bm{P}(k)\right\| _2 ^2\right) \\
    &-\frac{\rho}{2}\sum\limits_{k=1}^K\left(\left\|  \bm{P}(k)\right\| _2 ^2\right) \\
    \text{s.t.}\quad \mathbf{w}^{\rm H}&(k)\bm{a}(k)=1,\quad\enspace \text{for}~k=1,...,K,
\end{aligned}
\end{equation}
where $\rho > 0$ is the penalty parameter and $\bm{P}(k)$ denotes the scaled dual variable. Subsequently, we add the second constraint from MVDR to get the final Lagrange function:
\begin{equation}
\label{L2}
\small
\begin{aligned}
    \mathcal{L}( {\mathbf{w}}(k),&\bm{R},\bm{V},\bm{P}) =\frac{1}{N}\sum\limits_{k=1}^K\left|\left|  \mathbf{w}^{\rm H}(k){\bm{X}}(k)\right|\right| _2 ^2+\frac{\alpha}{2}\langle \,\bm{R}, \bm{R} - \Omega(\bm{R}) \,\rangle\\
    &+\frac{\rho}{2}\sum\limits_{k=1}^K\left(\left\|\mathbf{w}^{\rm H}(k){\bm{X}}(k)-\bm{R}(k)-\bm{V}(k)+\bm{P}(k)\right\| _2 ^2\right) \\
    &-\frac{\rho}{2}\sum\limits_{k=1}^K\left(\left\|  \bm{P}(k)\right\| _2 ^2\right) +\sum\limits_{k=1}^K{\bm \lambda}(k)\left(\mathbf{w}^{\rm H}(k)\bm{a}(k)-1\right),
\end{aligned}
\end{equation}

Rather than applying a unified optimization procedure, this two-step approach not only reduces computational complexity but also facilitates the convergence of the algorithm in subsequent iterations. By decoupling the optimization of equation~(\ref{L2}) into multiple subproblems, the variables can be updated iteratively across iterations indexed by $\ell$.

\vspace{0.15cm}
\noindent
\textbf{Step 1 }--- \textit{Optimization with respect to $\mathbf{w}(k) $:} The optimization of (\ref{L2}) reduces to:
\begin{equation}\small
    \begin{aligned}
        \mathbf{w}^{(\ell+1)}(k&)=\mathop{\rm argmin}\limits_{\mathbf{w}(k)}\frac{1}{N}\sum\limits_{k=1}^K\left\|  \mathbf{w}^{\rm H}(k){\bm{X}}(k)\right\| _2 ^2\\
        &+\frac{\rho}{2}\sum\limits_{k=1}^K\left(\left\|\mathbf{w}^{\rm H}(k){\bm{X}}(k)-\bm{R}(k)-\bm{V}(k)+\bm{P}(k)\right\| _2 ^2\right).\\
        &+\sum\limits_{k=1}^K({\bm \lambda(k)}\cdot(\mathbf{w}^{\rm H}(k)\cdot\bm{a}(k)-1)) \\
    \end{aligned}
\end{equation}
Setting the partial derivatives of $\mathbf{w}(k)$ and $\bm \lambda(k)$ to zero respectively, we could calculate the analytical solution of $\mathbf{w}(k)$ as the following form
\begin{equation}\label{w}
    \begin{aligned}
        \mathbf{w}^{(\ell+1)}(k)=\frac{1}{N\rho+2}\bm{R}_x^{-1}(k) \left\lbrace   \rho[\bm{Z}^{(\ell)}(k)]^{\rm H}-\bm{F}^{(\ell)}(k)\right\rbrace ,
    \end{aligned}
\end{equation}
where $\bm{R}_x(k)=\frac{1}{N}{\bm{X}}(k){\bm{X}}^{\rm H}(k)$ represents the covariance matrix of observations, with
\begin{equation}\small
    \bm{Z}^{(\ell+1)}(k)=\left(\bm{R}^{(\ell)}(k) + \bm{V}^{(\ell)}(k) - \bm{P}^{(\ell)}(k)\right){\bm{X}}^{\rm H}(k),
\end{equation}
and
\begin{equation}
    \bm{F}^{(\ell+1)}(k)=\frac{\rho\bm{Z}^{(\ell+1)}(k)\bm{R}_x^{-1}(k)\bm{a}^{(\ell)}(k)-\left( N\rho+2\right) }{[\bm{a}^{(\ell)}(k)]^{\rm H}\bm{R}_x^{-1}(k)\bm{a}^{(\ell)}(k)}\bm{a}^{(\ell)}(k).
\end{equation}
Subsequently, $\widehat{\bm S}$ is given by calculating
\begin{equation}\label{hats}
    \widehat{\bm{S}}^{(l+1)}(k) =[\mathbf{w}^{(\ell+1)}(k)]^{\rm H}{\bm{X}}(k)
\end{equation}
for each band.

\vspace{0.15cm}
\noindent \textbf{Step 2} --- \textit{Optimization with respect to $\bm R$:} The optimization problem (\ref{L2}) becomes
\begin{equation}
    \begin{aligned}
        \bm{R}^{(\ell+1)}&=\mathop{\rm argmin}\limits_{\bm{R}}\frac{\rho}{2}\left\|\widehat{\bm{S}}^{(\ell+1)}-\bm{R}^{(\ell)}-\bm{V}^{(\ell)}+\bm{P}^{(\ell)}\right\| _2 ^2\\
        & +\frac{\alpha}{2}\langle \,\bm{R}, \bm{R} - \Omega(\bm{R}) \,\rangle
    \end{aligned}
\end{equation}
From the perspective of RED, the speech prior can be incorporated in~\cite{red} by applying a denoising processing to speech $\widehat{\bm{R}}$ defined by
\begin{equation}
    \widehat{\bm{R}}^{(\ell+1)}=\widehat{\bm{S}}^{(\ell+1)}-\bm{V}^{(\ell)}+\bm{P}^{(\ell)}.
    \label{R_hat}
\end{equation}
Setting the derivative of the above optimization w.r.t. $\bm{R}$ to zero, we can minimize it directly by some iterative scheme, leading to the following equation:
\begin{equation}
    \rho\left(\bm R^{(\ell+1)} - \widehat{\bm R}^{(\ell+1)}\right) + \alpha\left(\bm R^{(\ell+1)} - \Omega(\bm R^{(\ell+1)})\right) = 0.
\end{equation}
From~\cite{Cohen_Elad_Milanfar_2021} and~\cite{yang2023integrating}, 
we can solve this problem via the fixed-point iteration leading to:
\begin{equation}\label{R_update}
    \widehat{\bm{R}}^{(\ell+1,i+1)}=\mu\widehat{\bm{R}}^{(\ell+1,i)}+(1-\mu)\Omega(\widehat{\bm{R}}^{(\ell+1,i)};\mathbf{\Theta}),
\end{equation}
with inner iteration $i=1,...,I$, where $\mu = \frac{\rho}{\rho+\alpha}$ is the trade-off parameter of the fixed-point method and $\mathbf{\Theta}$ denotes the learnable parameters of denoiser. The variable $ \bm{R}^{(\ell+1)}$ is equal to the final output of the fixed-point iteration.

\vspace{0.15cm}
\noindent \textbf{Step 3} --- \textit{Update of $\bm P$:} Following the dual variable update rule of the augmented Lagrangian, $ \bm{P} $ is updated in the standard manner:
\begin{equation}\label{P}
    \bm{P}^{(\ell+1)}=\bm{P}^{(\ell)}+\widehat{\bm{S}}^{(\ell+1)}-\bm{R}^{(\ell+1)}-\bm{V}^{(\ell)}.
\end{equation}

\vspace{0.15cm}
\noindent
\textbf{Step 4 }--- \textit{Optimization with respect to $\bm V$:} Finally, we set the derivative of (\ref{L2}) to zero as above to update the variable $ \bm{V}$:
\begin{equation}\label{V}
    \bm{V}^{(\ell+1)}=\widehat{\bm{S}}^{(\ell+1)}-\bm{R}^{(\ell+1)}+\bm{P}^{(\ell+1)}.
\end{equation}


The variables above are updated iteratively until convergence and the output $\bm R$ is regarded as the estimated target speech. The proposed PnP-MVDR algorithm is summarized in Algorithm \ref{alg:pnp}.

\begin{algorithm}[t!]
\caption{The proposed PnP-MVDR framework.}
\label{alg:pnp}
\KwIn {The observed multi-channel speech:~${\bm{X}}$}
\KwOut { The predicted speech:~$\bm R$}
\textbf{Parameters:}~The inner iterations:~$I$; The penalty parameter of the variable splitting method:~$\rho$; The parameter of the fixed-point algorithm: $\mu$\\
\textbf{Initialize:}   The auxiliary variables:~$\mathbf{V} = \boldsymbol{0}_{N\times K} $; The dual variables:~$\mathbf{P} =  \boldsymbol{0}_{N\times K} $; The predicted filter weight vector:~${\mathbf{w}}(k)= \boldsymbol{0}_{Q\times 1 } $.\\
\While{ not converged}{
    \For{k =1 to K}{
        Calculate $ {\mathbf{w}}^{(\ell+1)}(k)$ by (\ref{w}) \;
    }
    Calculate $\widehat{\bm S}^{(\ell+1)}$ by (\ref{hats}) \;
    Calculate $\bm R^{(\ell+1)}$ by (\ref{R_update}) \;
    Calculate $\bm P^{(\ell+1)}$ by (\ref{P}) \;
    Calculate $\bm V^{(\ell+1)}$ by (\ref{V}) \;
    $\ell = \ell+1$ }
\end{algorithm}

\section{Microphone Beamforming System and Real-world Implementation}
This section introduces our microphone beamforming system and describes its implementation in real-world scenarios.

\begin{figure*}
    \centering
    {
    \includegraphics[width=0.98\textwidth, trim=4.2cm 11.5cm 3.55cm 3.8cm, clip]{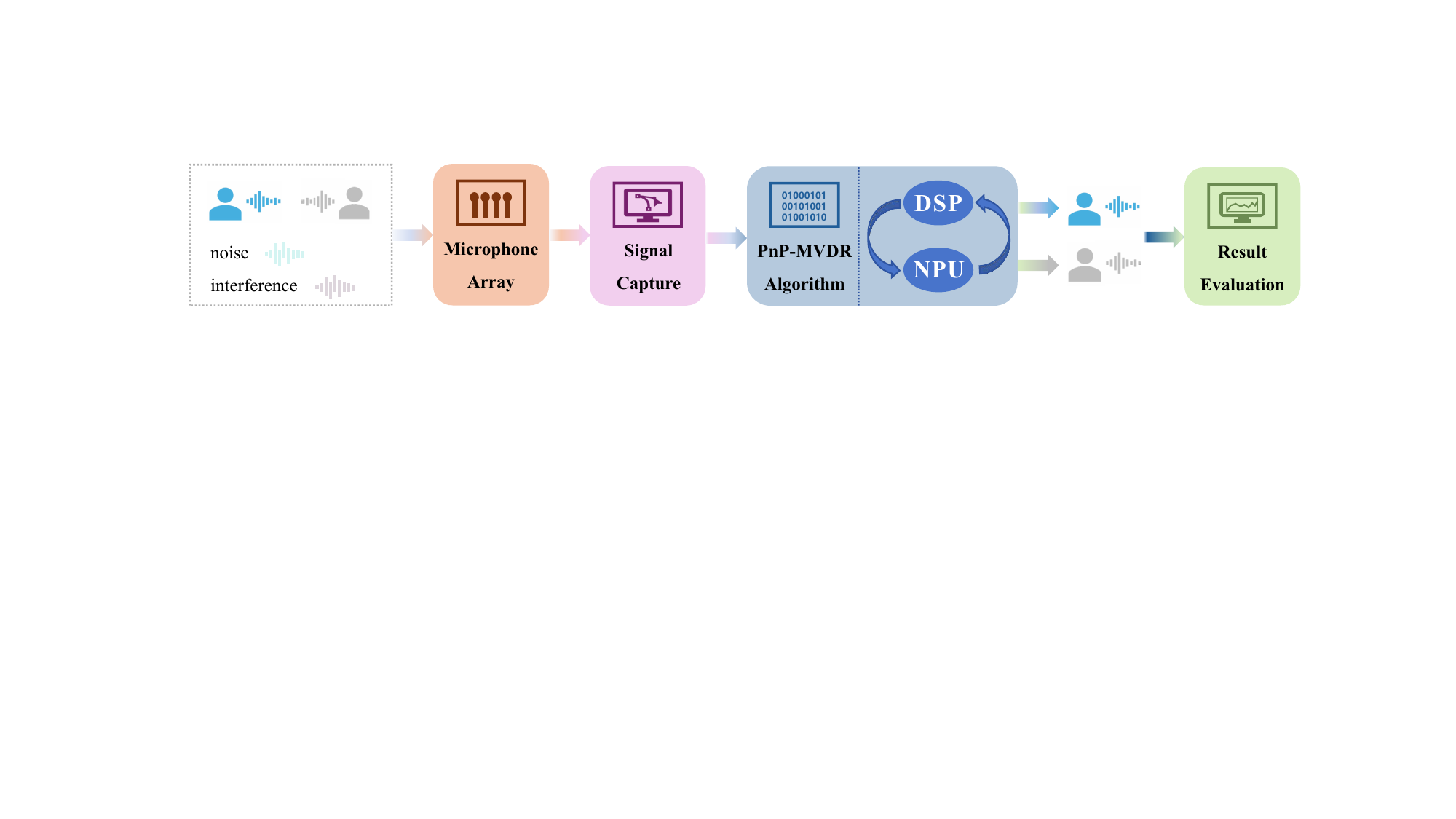}
    }
    \caption{ {Workflow of the proposed microphone beamforming system. The system captures mixed speech signals, which are then processed by a cooperative DSP-NPU architecture within the separation algorithm module to produce enhanced speech for evaluation. }}
    \label{fig:system}
\end{figure*}
\begin{figure*}
    \centering
    {
    \includegraphics[width=0.98\textwidth, trim = 0cm 0cm 1cm 0cm, clip]{{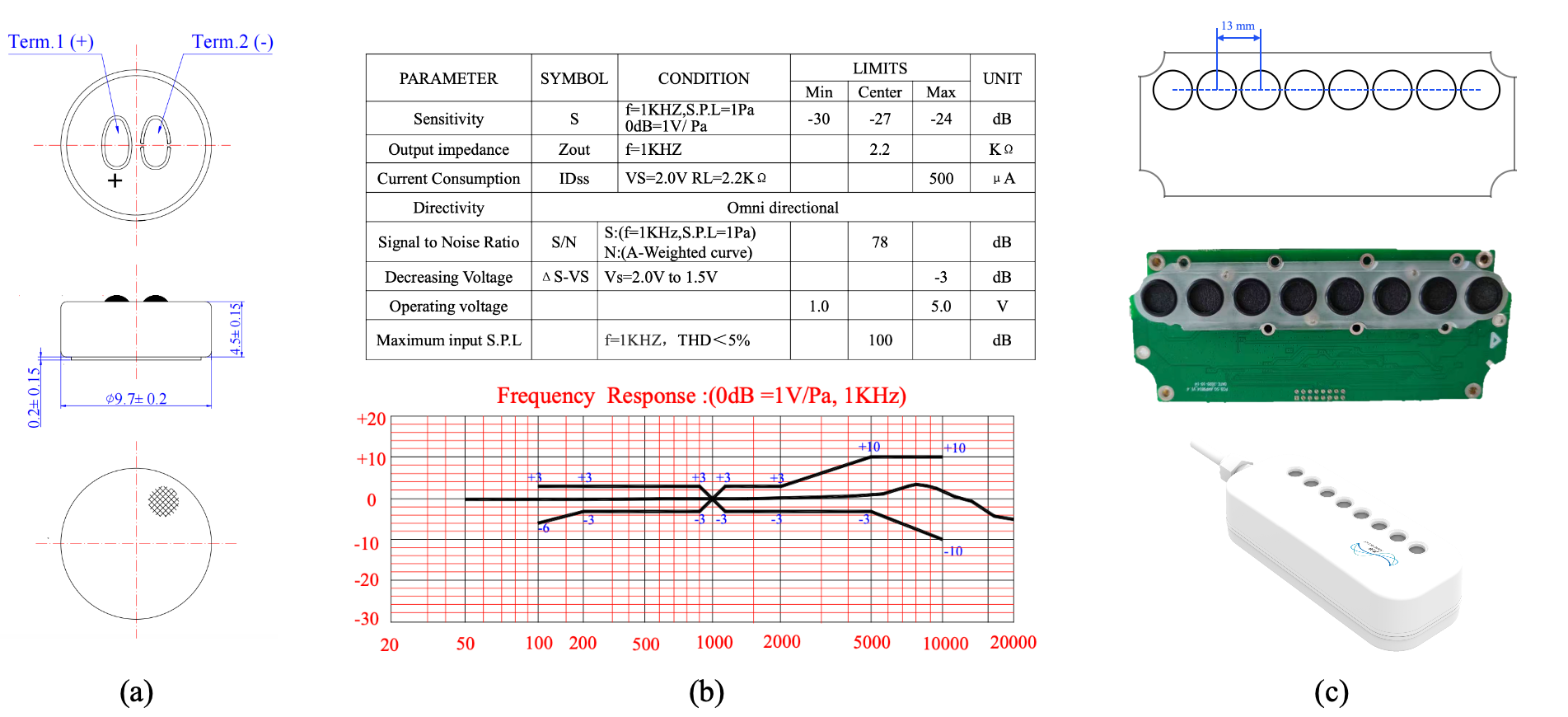}}
    }
    \caption{Detailed configuration of the microphone array. (a)The geometric dimensions of the microphone head, encompassing the top view, side view, and bottom view. (b)The technical specifications and frequency response of the microphone head. (c)The physical appearance of the microphone array.}
    \label{fig:mic}
\end{figure*}
\subsection{Microphone Beamforming System Analysis}

The overall architecture of our microphone beamforming system is illustrated in Fig.~\ref{fig:system}, which outlines the complete signal processing pipeline from acoustic capture to result evaluation. The system employs an 8-element linear microphone array that receives mixed acoustic signals containing target speech, interference, and background noise. These signals are transmitted via wired connection to the processing unit where analog-to-digital conversion and PCM encoding are performed. The core processing stage implements a cooperative DSP-NPU architecture for our PnP-MVDR framework. The DSP module handles real-time spatial covariance matrix updates and beamforming operations, while the NPU accelerates the DNN-based denoiser, which delivers superior speech enhancement performance.

From a computational perspective, the system demonstrates practical feasibility for actual deployment, as its computational demands are well-bounded. For example, a denoiser like FRCRN\cite{frcrn} requires approximately 4 Giga MAC operations per second, while the spatial processing stage with 512-point FFT maintains favorable computational characteristics. This level of demand aligns well with the 5-50 GOPS capacity of contemporary audio processing platforms, ensuring the practical implementability of the complete system pipeline as illustrated in Fig. \ref{fig:system}. Otherwise, the architectural design enables optimized workload distribution: the DSP handles signal acquisition and spatial feature extraction, generating high-resolution spectral representations that are asynchronously processed by the NPU for denoising inference. This partitioning strategy supports concurrent processing that significantly reduces end-to-end latency, while enabling intelligent power management—automatically transitioning between power-optimized modes for continuous listening and high-performance modes during active speech periods.

Following the algorithm module, the system outputs processed speech signals to the result evaluation module where algorithm performance is assessed using objective quality metrics. The integration of 512-point FFT processing with the sophisticated denoiser represents an optimal balance between traditional beamforming advantages and advanced deep learning capabilities, ensuring robust performance across various real-world acoustic environments while maintaining computational efficiency for practical deployment.

\subsection{Microphone Array Design}

To support the signal processing pipeline described above, we have developed a custom-designed microphone array that forms the front-end of our system. As shown in Fig.~\ref{fig:mic}, our group (CIAIC group of Northwestern Polytechnical University) has specifically designed this array to accurately capture complex acoustic signals and effectively extract spatial information, thereby ensuring the performance and stability of the complete system.

Each microphone head in the array has dual connection terminals, with compact dimensions of 9.7 mm (diameter) × 4.5 mm (height) to balance acoustic performance and spatial efficiency. A mesh is placed at the bottom of the microphone head to protect the internal sensor from particulate contamination.
The technical specifications and frequency response of the microphone head is shown in Fig.\ref{fig:mic} (b). This microphone sensor exhibits superior performance across multiple key metrics. It features a high-performance capsule designed for optimal sound capture across a wide frequency range. With a sensitivity between -30 dB and -24 dB at 1 kHz, it ensures clear audio reproduction, making it suitable for various acoustic environments. The omnidirectional directivity allows for effective sound capture from all angles, enhancing its versatility in recording settings. Additionally, the microphone has a signal-to-noise ratio of 78 dB, providing excellent audio clarity by reducing background noise interference. Its ability to handle a maximum input sound pressure level of 100 dB ensures durability and reliability, even in loud environments. 

Considering convenience, we choose to arrange the eight microphone heads in a linear, equidistant manner to collect multi-channel speech signals and further extract spatial information.
The spacing is set to 1.3 cm, which can effectively avoid spatial aliasing of high-frequency signals while improving high-frequency resolution.
Finally, the microphone array is housed in a white plastic enclosure, with its terminal side connected to the device via a cable as shown in Fig.\ref{fig:mic} (c).

This well-designed microphone array serves as the fundamental component of our system depicted in Fig.~\ref{fig:system}. It captures acoustic mixtures containing speech from multiple speakers, interference, and background noise, which are then transmitted to the processing unit through wired connection. The device software performs A/D conversion on the signal and encodes it into PCM audio format. Subsequently, the algorithm module processes the input and obtains the processed speech signal, and finally the performance of the algorithm is evaluated by the result evaluation module.

\subsection{Real-world Implementation}

\begin{figure*}
    \centering
    \includegraphics[width=0.98\textwidth, trim=2.3cm 1.7cm 2.2cm 2.1cm, clip]{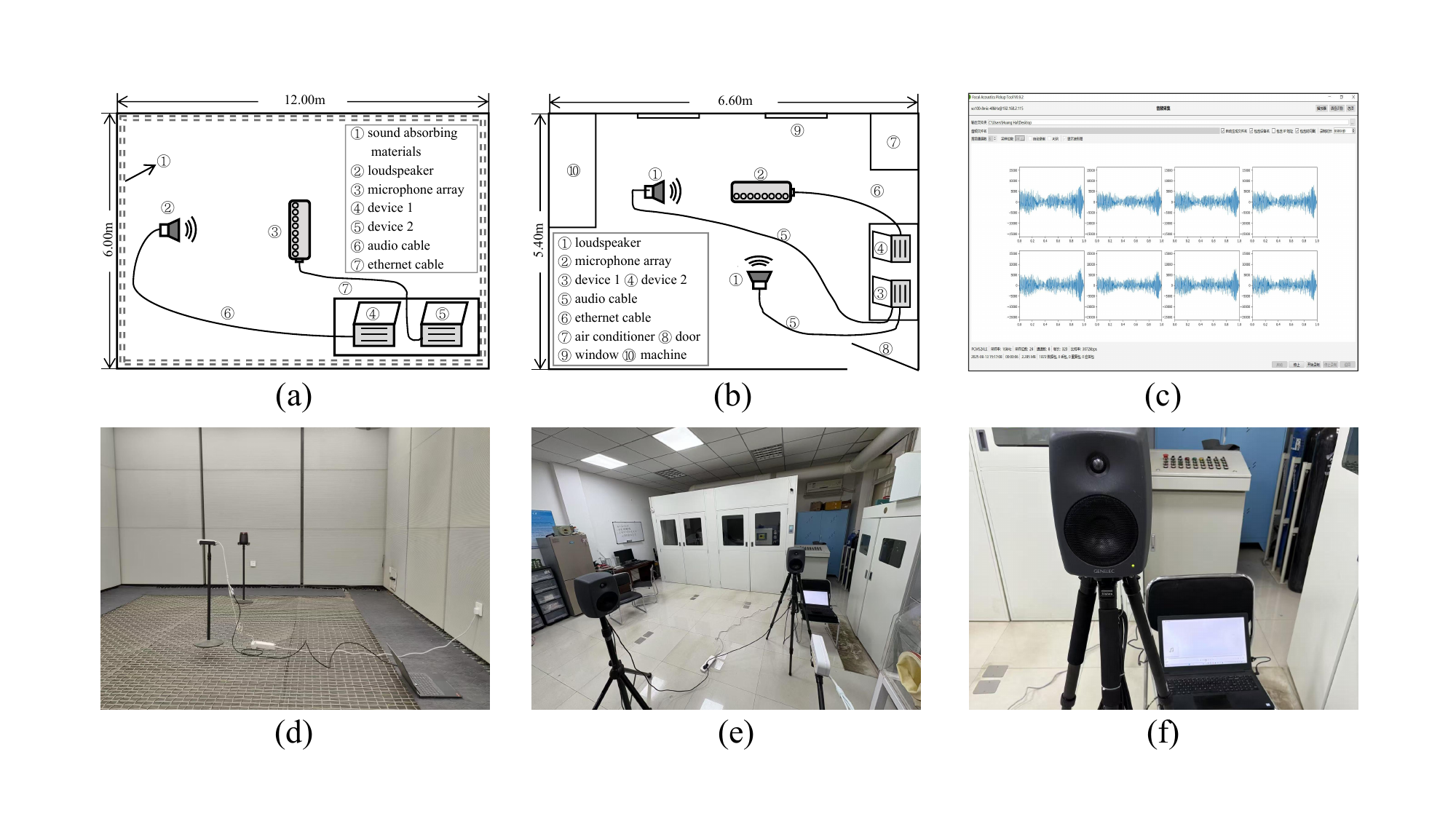}
    
    \caption{Experimental scenario diagram of capturing real-world data. 
    (a) Schematic diagram of the experimental setup in anechoic room where we capture clean speech signals for evaluation. (b) Schematic diagram of the experimental setup in reverberant enclosed room where we record the received speech signals for test. (c) The signal acquisition software and interface associated to the microphone array. (d)The physical implementation diagram corresponding to (a). (e)The physical implementation diagram corresponding to (b). (f)The Canon loudspeaker we used to play the speech signal. }
    \label{fig:real-world}
\end{figure*}

To assess the practical feasibility of the custom-designed microphone beamforming system, we employ it to capture real-world data as illustrated in Fig.~\ref{fig:real-world}. Specifically, clean speech signals are collected in an anechoic room, while input speech signals are recorded in a reverberant enclosed room. Fig.~\ref{fig:real-world}(a) and (b) present schematic diagrams of the experimental setups in the anechoic and reverberant rooms, respectively, and Fig.~\ref{fig:real-world}(c) shows the playback and recording devices used for audio acquisition.

In the experimental procedure, a loudspeaker is placed in the anechoic room to play designated speech signals at a distance from the microphone array, providing clean reference signals for subsequent evaluation, as depicted in Fig.~\ref{fig:real-world}(d). In the reverberant enclosed room, the microphone array and two Canon loudspeakers are used simultaneously to play two clean speech signals, as shown in Fig.~\ref{fig:real-world}(e). The signals captured by the microphone array correspond to mixed speech, which serve as the input to the speech separation module. Additionally, Fig.~\ref{fig:real-world}(f) presents the Canon loudspeaker employed in the experiments.

\section{Experiment Setup}
This section describes the experimental settings, including the datasets, comparison methods, evaluation metrics, and experimental details.

\subsection{Datasets}
In order to validate the superiority of the proposed method, we conduct tests on both simulated data and real-world data. 

\textit{1) Simulated Data:}
we generate mixed speech signals and clean speech signals using the image method in~\cite{rir}. Specifically, we first generate a room with its length, width, and height set as 4, 4 and 3 meters, respectively. The reverberation time T60 in the room is set to 0. And then adopts an 8-element linear microphone array to receive the speech signals with an array spacing of 0.03m. The two speech sources are placed randomly in the room and it is certain that the distance between them was controlled to be in [0.5, 3] meters. Note that each individual source signal received by the reference array element is used as the corresponding clean signal. Subsequently, we split 10 pairs of clean speech signals, randomly chosen from the Wall Street Journal dataset (WSJ0, \cite{wsj0}) and convolve them with the RIRs. 

In addition, we evaluate the robustness of the proposed framework under various noise conditions. As widely demonstrated in image processing based on PnP, this strategy to excavate deep priors is able to yield promising results with White Gaussian Noise (WGN). However, for speech signal processing, it is insufficient to test the performance only in the case of WGN, and thus we consider more scenarios. In our experiments, we choose three additive noises, namely WGN, babble, and cafe, chosen from the diffuse noise set, and further add them to the mixed speech signals. We set the signal-to-noise ratios (SNR) to 0 dB, 10 dB, and 20 dB, respectively. In this way, we obtain a test set of 9 experimental scenarios and a total of 90 mixed utterances. 

\textit{2) Real-world Data:}
To test our proposed system and validate the robustness of the core speech separation algorithm under real scenarios, we also collect real-world data.
Firstly, we select 10 pairs of speech signals, randomly chosen from WSJ0. And then, these speech signals are played and collected in anechoic room to obtain clean speech signal. Finally, each pair of speech signals is played through two Canon loudspeakers in the reverberant enclosed room and received by microphone array. 
Furthermore, we also introduce three types of additive interference: \textit{cafe}, \textit{street}, and \textit{factory} noise, selected from standard diffuse noise corpora,  to assess the robustness of the proposed algorithm under realistic noise conditions.
In this way, we obtain a test set of 3 experimental scenarios and a total of 30 mixed utterances. 

\subsection{Experimental configurations}
Following dataset preparation, the data undergoes essential preprocessing steps. For simulated data, each utterance is split into 6.4 seconds with a sampling rate of 16 kHz before implementing the separation algorithm. For real-world data, we first conduct algorithm processing to obtain the two separated signals, which will be aligned with the corresponding clean signal using the gcc-phat function of MATLAB, and then truncated into 6.4 seconds to evaluate and compute the metrics. In addition, the simulation data and real data adopt the same STFT configuration, specifically, a 512-point FFT is applied to time-domain signal along with 32 ms Hann window and 16 ms step size.

In our framework, there are several parameters that need to be adjusted, including the inner iteration $I$, the outer iteration $N$, the penalty parameter $ \rho$ and the trade-off parameter $\mu$. For simplicty, we set the inner iteration $I$ to 1 to speed the overall framework. The penalty parameter $\rho$ in the augmented Lagrangian function only affects the convergence speed, as theoretically shown~\cite{zhao2021plug}. Therefore we simply set its value from \{0.1, 1, 10\} in our experiments. The trade-off parameter $\mu$ in the fixed-point iteration plays a critical role in controlling the influence of the denoising engine. Specifically, if $\mu$ is set excessively high, the intermediate variable $\hat{\bm{R}}$ in [\ref{R_hat}] becomes predominant, resulting in a less effective denoiser. Conversely, setting $\mu$ to a low value enhances the denoiser's strength, while diminishes the separation result related to the optimization problem w.r.t. $w(k)$. Experientially, we first choose a relatively suitable setting on a single signal for each experimental scenario, and then apply the selected setting to all test sets.

To mitigate randomness, the methods are evaluated on the test set, and the average of all results is calculated to demonstrate performance.

\subsection{Comparison method}
As mentioned above, an off-the-shelf denoiser is utilized to incorporate speech prior information into our framework. In addition, benefiting from the generalization ability of the proposed method, any speech-applicable denoiser can be employed, and also the ability of the denoiser to learn unique characteristics from speech data directly affects the performance of our framework. On the other hand,  it shows that our framework is quite flexible and can be improved along with the improvement of the denoiser.
\begin{table}[t]
\caption{Details about DNN-based denoisers which are plugged into the proposed framework}
\label{denoiser}
\centering
\renewcommand{\arraystretch}{1.5}
\begin{tabular}{c|c|c}
\hline
DNN-based denoiser                       & Training dataset                            &  Neural network structure            \\ \hline
DCUNet \cite{dcunet}    & Libri1Mix \cite{libri1mix} &  Deep Complex U-Net \\\hline
DCCRNet \cite{dccrnet}  & Libri1Mix                                   & Deep Complex CRN                    \\\hline
FRCRN \cite{frcrn}      & DNS-Challenge \cite{dns}   &  Frequency Recurret CRN              \\ \hline
\end{tabular}
\end{table}

\begin{table*}[t!]
\centering
\caption{The simulated experimental results of all comparison methods under different noise levels. The best results are in bold.}
\renewcommand\arraystretch{1.35}
\label{simulated}
\begin{tabular}{c|c|c|ccc|ccc|ccc}
\toprule
\multirow{4}{*}{Noise type} & \multicolumn{2}{c|}{\multirow{2}{*}{Noise level}}& \multicolumn{3}{c|}{\multirow{2}{*}{SNR = 20dB}} & \multicolumn{3}{c|}{\multirow{2}{*}{SNR = 10dB}} & \multicolumn{3}{c}{\multirow{2}{*}{SNR = 0dB}} \\
&\multicolumn{2}{c|}{}& \multicolumn{3}{c|}{} & \multicolumn{3}{c|}{} & \multicolumn{3}{c}{} \\ \cline{2-12}& {\multirow{2}{*}{Method}} &\multirow{2}{*}{Denoise}& \multicolumn{1}{c|}{\multirow{2}{*}{fwSNRseg}} & \multicolumn{1}{c|}{\multirow{2}{*}{PESQ}} & \multirow{2}{*}{STOI} & \multicolumn{1}{c|}{\multirow{2}{*}{fwSNRseg}} & \multicolumn{1}{c|}{\multirow{2}{*}{PESQ}} & \multirow{2}{*}{STOI} & \multicolumn{1}{c|}{\multirow{2}{*}{fwSNRseg}} & \multicolumn{1}{c|}{\multirow{2}{*}{PESQ}} & \multirow{2}{*}{STOI} \\
&   && \multicolumn{1}{c|}{} & \multicolumn{1}{c|}{} &  & \multicolumn{1}{c|}{} & \multicolumn{1}{c|}{} &  & \multicolumn{1}{c|}{} & \multicolumn{1}{c|}{} &  \\ \hline
\multirow{13}{*}{WGN} & Unprocessed  &-& \multicolumn{1}{c|}{8.13} & \multicolumn{1}{c|}{1.15} & 0.74 & \multicolumn{1}{c|}{6.29} & \multicolumn{1}{c|}{1.08} & 0.72 & \multicolumn{1}{c|}{5.13} & \multicolumn{1}{c|}{1.05} & 0.63 \\ \cline{2-12}& MVDR  &-& \multicolumn{1}{c|}{14.39} & \multicolumn{1}{c|}{1.92} & 0.99 & \multicolumn{1}{c|}{10.34} & \multicolumn{1}{c|}{1.30} & 0.97 & \multicolumn{1}{c|}{7.13} & \multicolumn{1}{c|}{1.09} & 0.89 \\ \cline{2-12}& PIT  &-& \multicolumn{1}{c|}{7.77} & \multicolumn{1}{c|}{1.49} & 0.89 & \multicolumn{1}{c|}{5.92} & \multicolumn{1}{c|}{1.30} & 0.86 & \multicolumn{1}{c|}{3.04} & \multicolumn{1}{c|}{1.11} & 0.69 \\ \cline{2-12}& Sepformer  &-& \multicolumn{1}{c|}{10.91} & \multicolumn{1}{c|}{1.80} & 0.96 & \multicolumn{1}{c|}{9.82} & \multicolumn{1}{c|}{1.67} & 0.95 & \multicolumn{1}{c|}{7.96} & \multicolumn{1}{c|}{1.28} & 0.89 \\ \cline{2-12}& DeepMVDR  &-& \multicolumn{1}{c|}{9.00} & \multicolumn{1}{c|}{1.85} & 0.90 & \multicolumn{1}{c|}{7.71} & \multicolumn{1}{c|}{1.69} & 0.89 & \multicolumn{1}{c|}{5.40} & \multicolumn{1}{c|}{1.26} & 0.82 \\ \cline{2-12}
& \multirow{3}{*}{Denoise+MVDR}&DCCRNet& \multicolumn{1}{c|}{10.1} & \multicolumn{1}{c|}{1.54} & 0.91& \multicolumn{1}{c|}{9.02} & \multicolumn{1}{c|}{1.32} & 0.88& \multicolumn{1}{c|}{7.30} & \multicolumn{1}{c|}{1.12} & 0.76\\ \cline{3-12}
& &DCUNet& \multicolumn{1}{c|}{12.13 } & \multicolumn{1}{c|}{1.83 } & 0.95 
& \multicolumn{1}{c|}{10.09} & \multicolumn{1}{c|}{1.50} & 0.88 & \multicolumn{1}{c|}{7.77} & \multicolumn{1}{c|}{1.16} & 0.83\\\cline{3-12}&  &FRCRN& \multicolumn{1}{c|}{10.12} & \multicolumn{1}{c|}{1.53} & 0.83 & \multicolumn{1}{c|}{8.74} & \multicolumn{1}{c|}{1.55} & 0.88 & \multicolumn{1}{c|}{6.08} & \multicolumn{1}{c|}{1.34} & 0.82 \\ \cline{2-12}& \multirow{3}{*}{MVDR+Denoise}  &DCCRNet& \multicolumn{1}{c|}{13.73} & \multicolumn{1}{c|}{2.08} & 0.99 & \multicolumn{1}{c|}{12.28} & \multicolumn{1}{c|}{1.71} & 0.98 & \multicolumn{1}{c|}{8.80} & \multicolumn{1}{c|}{1.24} & 0.93 \\ \cline{3-12}&    &DCUNet& \multicolumn{1}{c|}{15.14} & \multicolumn{1}{c|}{2.52} & \textbf{1.00} & \multicolumn{1}{c|}{12.78} & \multicolumn{1}{c|}{2.05} & 0.98 & \multicolumn{1}{c|}{9.31} & \multicolumn{1}{c|}{1.34} & 0.94 \\ \cline{3-12}&   &FRCRN& \multicolumn{1}{c|}{15.53} & \multicolumn{1}{c|}{\textbf{3.03}} & \textbf{1.00} & \multicolumn{1}{c|}{13.01} & \multicolumn{1}{c|}{\textbf{2.75}} & \textbf{0.99} & \multicolumn{1}{c|}{10.50} & \multicolumn{1}{c|}{1.89} & \textbf{0.97} \\ \cline{2-12}& \multirow{3}{*}{\cellcolor{gray!30}PnP-MVDR}  &\cellcolor{gray!30}DCCRNet
& \multicolumn{1}{c|}{\cellcolor{gray!30}17.52} & \multicolumn{1}{c|}{\cellcolor{gray!30}2.43} & \cellcolor{gray!30}0.99 & \multicolumn{1}{c|}{\cellcolor{gray!30}14.53} & \multicolumn{1}{c|}{\cellcolor{gray!30}1.98} & \cellcolor{gray!30}0.98 & \multicolumn{1}{c|}{\cellcolor{gray!30}10.19} & \multicolumn{1}{c|}{\cellcolor{gray!30}1.35} & \cellcolor{gray!30}0.94 \\ \cline{3-12}
& \cellcolor{gray!30}&\cellcolor{gray!30}DCUNet
& \multicolumn{1}{c|}{\cellcolor{gray!30}17.61} & \multicolumn{1}{c|}{\cellcolor{gray!30}2.66} & \cellcolor{gray!30}\textbf{1.00} & \multicolumn{1}{c|}{\cellcolor{gray!30}14.28} & \multicolumn{1}{c|}{\cellcolor{gray!30}2.18} & \cellcolor{gray!30}0.98 & \multicolumn{1}{c|}{\cellcolor{gray!30}9.67} & \multicolumn{1}{c|}{\cellcolor{gray!30}1.37} & \cellcolor{gray!30}0.95 \\ \cline{3-12}
&\cellcolor{gray!30} \multirow{-3}{*}{PnP-MVDR}&\cellcolor{gray!30}FRCRN& \multicolumn{1}{c|}{\cellcolor{gray!30}\textbf {18.23}} & \multicolumn{1}{c|}{\cellcolor{gray!30}2.92} & \cellcolor{gray!30}\textbf{1.00} & \multicolumn{1}{c|}{\cellcolor{gray!30}\textbf{15.34}} & \multicolumn{1}{c|}{\cellcolor{gray!30}2.54} & \cellcolor{gray!30}\textbf{0.99} & \multicolumn{1}{c|}{\cellcolor{gray!30}\textbf{12.08} }& \multicolumn{1}{c|}{\cellcolor{gray!30}1.72} & \cellcolor{gray!30}\textbf{0.97} \\ \hline
\multirow{13}{*}{Cafe} & Unprocessed  &-
& \multicolumn{1}{c|}{9.00} & \multicolumn{1}{c|}{1.18} & 0.74 & \multicolumn{1}{c|}{6.62} & \multicolumn{1}{c|}{1.13} & 0.71 & \multicolumn{1}{c|}{4.46} & \multicolumn{1}{c|}{1.06} & 0.63 \\ \cline{2-12}& MVDR  &-
& \multicolumn{1}{c|}{15.07} & \multicolumn{1}{c|}{2.27} & 0.99 & \multicolumn{1}{c|}{12.26} & \multicolumn{1}{c|}{1.90} & 0.98 & \multicolumn{1}{c|}{10.34} & \multicolumn{1}{c|}{1.72} & 0.95 \\ \cline{2-12}& PIT  &-
& \multicolumn{1}{c|}{8.34} & \multicolumn{1}{c|}{1.50} & 0.89 & \multicolumn{1}{c|}{7.12} & \multicolumn{1}{c|}{1.35} & 0.85 & \multicolumn{1}{c|}{4.71} & \multicolumn{1}{c|}{1.11} & 0.71 \\ \cline{2-12}& Sepformer  &-
& \multicolumn{1}{c|}{11.78} & \multicolumn{1}{c|}{2.01} & 0.96 & \multicolumn{1}{c|}{11.08} & \multicolumn{1}{c|}{\textbf{1.89}} & 0.96 & \multicolumn{1}{c|}{9.47} & \multicolumn{1}{c|}{1.56} & 0.92 \\ \cline{2-12}& DeepMVDR  &-
& \multicolumn{1}{c|}{8.25} & \multicolumn{1}{c|}{1.73} & 0.90 & \multicolumn{1}{c|}{7.27} & \multicolumn{1}{c|}{1.58} & 0.88 & \multicolumn{1}{c|}{6.11} & \multicolumn{1}{c|}{1.29} & 0.84 \\\cline{2-12}
&\multirow{3}{*}{ Denoise+MVDR} &DCCRNet
& \multicolumn{1}{c|}{10.34} & \multicolumn{1}{c|}{1.63} & 0.91& \multicolumn{1}{c|}{9.26} & \multicolumn{1}{c|}{1.49} & 0.89& \multicolumn{1}{c|}{7.26} & \multicolumn{1}{c|}{1.35} & 0.83\\ \cline{3-12}
&    &DCUNet
& \multicolumn{1}{c|}{12.09} & \multicolumn{1}{c|}{1.84} & 0.93& \multicolumn{1}{c|}{9.88} & \multicolumn{1}{c|}{1.62} & 0.90& \multicolumn{1}{c|}{7.36} & \multicolumn{1}{c|}{1.40} & 0.83\\ \cline{3-12}
&    &FRCRN
& \multicolumn{1}{c|}{11.36} & \multicolumn{1}{c|}{1.46} & 0.79 & \multicolumn{1}{c|}{9.71} & \multicolumn{1}{c|}{1.39} & 0.82 & \multicolumn{1}{c|}{8.16} & \multicolumn{1}{c|}{1.35} & 0.84 \\ \cline{2-12}& \multirow{3}{*}{MVDR+Denoise}  &DCCRNet
& \multicolumn{1}{c|}{11.36} & \multicolumn{1}{c|}{2.36} & 0.99 & \multicolumn{1}{c|}{13.42} & \multicolumn{1}{c|}{2.26} & \textbf{0.99} & \multicolumn{1}{c|}{12.81} & \multicolumn{1}{c|}{2.09} & \textbf{0.98} \\ \cline{3-12}&    &DCUNet
& \multicolumn{1}{c|}{16.65} & \multicolumn{1}{c|}{2.79} & \textbf{1.00} & \multicolumn{1}{c|}{14.87} & \multicolumn{1}{c|}{2.57} &\textbf{0.99} & \multicolumn{1}{c|}{14.00} & \multicolumn{1}{c|}{2.41} & \textbf{0.98} \\ \cline{3-12}&    &FRCRN
& \multicolumn{1}{c|}{17.46} & \multicolumn{1}{c|}{3.13} & 0.99 & \multicolumn{1}{c|}{16.26} & \multicolumn{1}{c|}{2.83} & \textbf{0.99} & \multicolumn{1}{c|}{15.68} & \multicolumn{1}{c|}{2.59} & \textbf{0.98} \\ \cline{2-12}& \multirow{3}{*}{\cellcolor{gray!30}PnP-MVDR}  &\cellcolor{gray!30}DCCRNet
& \multicolumn{1}{c|}{\cellcolor{gray!30}16.75} & \multicolumn{1}{c|}{\cellcolor{gray!30}2.53} & \cellcolor{gray!30}0.99 & \multicolumn{1}{c|}{\cellcolor{gray!30}14.14} & \multicolumn{1}{c|}{\cellcolor{gray!30}2.42} & \cellcolor{gray!30}\textbf{0.99 }& \multicolumn{1}{c|}{\cellcolor{gray!30}13.10} & \multicolumn{1}{c|}{\cellcolor{gray!30}2.30} & \cellcolor{gray!30}\textbf{0.98} \\ \cline{3-12} 
&  \cellcolor{gray!30}  &\cellcolor{gray!30}
DCUNet
& \multicolumn{1}{c|}{\cellcolor{gray!30}16.93} & \multicolumn{1}{c|}{\cellcolor{gray!30}2.68} &\cellcolor{gray!30}0.99 & \multicolumn{1}{c|}{\cellcolor{gray!30}14.88} & \multicolumn{1}{c|}{\cellcolor{gray!30}2.66} & \cellcolor{gray!30}\textbf{0.99 }& \multicolumn{1}{c|}{\cellcolor{gray!30}14.16} & \multicolumn{1}{c|}{\cellcolor{gray!30}2.50} & \cellcolor{gray!30}\textbf{0.98} \\ \cline{3-12}&  \cellcolor{gray!30}  \multirow{-3}{*}{PnP-MVDR} &\cellcolor{gray!30}FRCRN& \multicolumn{1}{c|}{\cellcolor{gray!30}\textbf{18.72}} & \multicolumn{1}{c|}{\cellcolor{gray!30}\textbf{3.15}} & \cellcolor{gray!30}0.99 & \multicolumn{1}{c|}{\cellcolor{gray!30}\textbf{16.48}} & \multicolumn{1}{c|}{\cellcolor{gray!30}\textbf{3.07} }&\cellcolor{gray!30}\textbf{0.99}& \multicolumn{1}{c|}{\cellcolor{gray!30} \textbf{15.91}} & \multicolumn{1}{c|}{\cellcolor{gray!30}\textbf{2.76}} & \cellcolor{gray!30}\textbf{0.98} \\ \hline
\multirow{13}{*}{Babble} & Unprocessed  &
-
& \multicolumn{1}{c|}{8.55} & \multicolumn{1}{c|}{1.19} & 0.74 & \multicolumn{1}{c|}{6.15} & \multicolumn{1}{c|}{1.14} & 0.69 & \multicolumn{1}{c|}{3.97} & \multicolumn{1}{c|}{1.07} & 0.57 \\ \cline{2-12} 
& MVDR  &
-& \multicolumn{1}{c|}{14.11} & \multicolumn{1}{c|}{2.24} & 0.99 & \multicolumn{1}{c|}{11.73} & \multicolumn{1}{c|}{1.92} & 0.97 & \multicolumn{1}{c|}{10.24} & \multicolumn{1}{c|}{1.74} & 0.94 \\ \cline{2-12} 
& PIT  &-
& \multicolumn{1}{c|}{8.05} & \multicolumn{1}{c|}{1.50} & 0.88 & \multicolumn{1}{c|}{6.56} & \multicolumn{1}{c|}{1.33} & 0.82 & \multicolumn{1}{c|}{3.86} & \multicolumn{1}{c|}{1.11} & 0.59 \\ \cline{2-12} 
& Sepformer  &
-& \multicolumn{1}{c|}{11.55} & \multicolumn{1}{c|}{2.05} & 0.96 & \multicolumn{1}{c|}{10.44} & \multicolumn{1}{c|}{1.82} & 0.95 & \multicolumn{1}{c|}{7.88} & \multicolumn{1}{c|}{1.31} & 0.83 \\ \cline{2-12} 
& DeepMVDR  &-
& \multicolumn{1}{c|}{8.11} & \multicolumn{1}{c|}{1.77} & 0.89 & \multicolumn{1}{c|}{6.78} & \multicolumn{1}{c|}{1.61} & 0.86 & \multicolumn{1}{c|}{5.28} & \multicolumn{1}{c|}{1.31} & 0.80 \\ \cline{2-12} 
& \multirow{3}{*}{Denoise+MVDR}  &
DCCRNet& \multicolumn{1}{c|}{9.73} & \multicolumn{1}{c|}{1.58} & 0.91& \multicolumn{1}{c|}{8.09} & \multicolumn{1}{c|}{1.38} & 0.88 & \multicolumn{1}{c|}{5.42} & \multicolumn{1}{c|}{1.21} & 0.77 \\ \cline{3-12} 
&    &
DCUNet& \multicolumn{1}{c|}{11.61} & \multicolumn{1}{c|}{1.7} & 0.93 & \multicolumn{1}{c|}{9.07} & \multicolumn{1}{c|}{1.56} & 0.89 & \multicolumn{1}{c|}{5.88} & \multicolumn{1}{c|}{1.28} & 0.77 \\ \cline{3-12} 
&    &
FRCRN& \multicolumn{1}{c|}{9.58} & \multicolumn{1}{c|}{1.43} & 0.76 & \multicolumn{1}{c|}{7.91} & \multicolumn{1}{c|}{1.41} & 0.75 & \multicolumn{1}{c|}{5.51} & \multicolumn{1}{c|}{1.21} & 0.70 \\ \cline{2-12} 
& \multirow{3}{*}{MVDR+Denoise}  &DCCRNet& \multicolumn{1}{c|}{13.12} & \multicolumn{1}{c|}{2.30} & \textbf{0.99} & \multicolumn{1}{c|}{12.37} & \multicolumn{1}{c|}{2.12} & 0.98 & \multicolumn{1}{c|}{11.43} & \multicolumn{1}{c|}{2.03} & \textbf{0.98} \\ \cline{3-12} 
&   &DCUNet
& \multicolumn{1}{c|}{15.11} & \multicolumn{1}{c|}{2.68} &\textbf{0.99} & \multicolumn{1}{c|}{13.88} & \multicolumn{1}{c|}{2.42} & 0.98 & \multicolumn{1}{c|}{12.58} & \multicolumn{1}{c|}{2.35} & \textbf{0.98} \\ \cline{3-12} 
&   &FRCRN
& \multicolumn{1}{c|}{16.41} & \multicolumn{1}{c|}{\textbf{3.20}} & \textbf{0.99} & \multicolumn{1}{c|}{15.30} & \multicolumn{1}{c|}{2.80} & \textbf{0.99} & \multicolumn{1}{c|}{15.80} & \multicolumn{1}{c|}{2.76} & \textbf{0.98} \\ \cline{2-12} 
& \cellcolor{gray!30}  &\cellcolor{gray!30}DCCRNet
& \multicolumn{1}{c|}{\cellcolor{gray!30}15.85} & \multicolumn{1}{c|}{\cellcolor{gray!30}2.56} & \cellcolor{gray!30}\textbf{0.99} & \multicolumn{1}{c|}{\cellcolor{gray!30}13.44} & \multicolumn{1}{c|}{\cellcolor{gray!30}2.63} & \cellcolor{gray!30}\textbf{0.99} & \multicolumn{1}{c|}{\cellcolor{gray!30}12.23} & \multicolumn{1}{c|}{\cellcolor{gray!30}2.33} & \cellcolor{gray!30}\textbf{0.98} \\ \cline{3-12}  
&  \cellcolor{gray!30}  &\cellcolor{gray!30}DCUNet
& \multicolumn{1}{c|}{\cellcolor{gray!30}16.00} & \multicolumn{1}{c|}{\cellcolor{gray!30}2.72} &\cellcolor{gray!30}\textbf{0.99} & \multicolumn{1}{c|}{\cellcolor{gray!30}14.31} & \multicolumn{1}{c|}{\cellcolor{gray!30}2.65} & \cellcolor{gray!30}\textbf{0.99} & \multicolumn{1}{c|}{\cellcolor{gray!30}12.29} & \multicolumn{1}{c|}{\cellcolor{gray!30}2.48} & \cellcolor{gray!30}\textbf{0.98} \\ \cline{3-12}  
&  \cellcolor{gray!30} \multirow{-3}{*}{PnP-MVDR} &\cellcolor{gray!30}FRCRN
& \multicolumn{1}{c|}{\cellcolor{gray!30}\textbf{17.66}} & \multicolumn{1}{c|}{\cellcolor{gray!30}3.14} &\cellcolor{gray!30}\textbf{0.99} & \multicolumn{1}{c|}{\cellcolor{gray!30}\textbf{15.71}} & \multicolumn{1}{c|}{\cellcolor{gray!30}\textbf{2.88}} & \cellcolor{gray!30}\textbf{0.99} & \multicolumn{1}{c|}{\cellcolor{gray!30}\textbf{15.90}} & \multicolumn{1}{c|}{\cellcolor{gray!30}\textbf{2.82}} & \cellcolor{gray!30}\textbf{0.98}\\
\bottomrule
\end{tabular}
\end{table*}
\begin{table}[t!]
\centering
\caption{The real-world experimental results of all comparison methods under different noise levels. The best results are in bold.}
\renewcommand\arraystretch{1.35}
\label{real-world}
\begin{tabular}{>{\centering\arraybackslash}p{0.7cm}|>{\centering\arraybackslash}p{1.85cm}|c|>{\centering\arraybackslash}p{0.5cm}cc}
\toprule
\multirow{1}{*}{Noise} & \multirow{2}{*}{Method}  &\multirow{2}{*}{Denoiser}& \multicolumn{3}{c}{Metrics} \\ \cline{4-6} 
\multirow{1}{*}{type}&   & & \multicolumn{1}{c|}{fwSNRseg} & \multicolumn{1}{c|}{PESQ} & STOI \\ \hline
\multirow{13}{*}{Cafe} & Unprocessed  &-& \multicolumn{1}{c|}{4.94} & \multicolumn{1}{c|}{1.16} &  0.55
\\ \cline{2-6} 
& MVDR  &-& \multicolumn{1}{c|}{5.20 } & \multicolumn{1}{c|}{1.23 } & 0.67 
\\ \cline{2-6} 
& PIT  &-& \multicolumn{1}{c|}{2.48} & \multicolumn{1}{c|}{1.17} &  0.56
\\ \cline{2-6}  & Sepformer  &-& \multicolumn{1}{c|}{6.38 } & \multicolumn{1}{c|}{1.31 } & 0.76 
\\ \cline{2-6} 
& DeepMVDR  &-& \multicolumn{1}{c|}{2.56} & \multicolumn{1}{c|}{1.19} &  0.60
\\ \cline{2-6} 
& \multirow{3}{*}{Denoise+MVDR } &DCCRNet& \multicolumn{1}{c|}{4.24 } & \multicolumn{1}{c|}{1.21 } & 0.44 
\\ \cline{3-6} 
&    &DCUNet& \multicolumn{1}{c|}{4.43 } & \multicolumn{1}{c|}{1.25 } & 0.44 
\\ \cline{3-6}
&    &FRCRN& \multicolumn{1}{c|}{5.37} & \multicolumn{1}{c|}{1.35 } & 0.45 
\\ \cline{2-6}
& \multirow{3}{*}{MVDR+Denoise}  &DCCRNet& \multicolumn{1}{c|}{5.23 } & \multicolumn{1}{c|}{1.29 } & 0.80 
\\ \cline{3-6} 
&   &DCUNet& \multicolumn{1}{c|}{5.12 } & \multicolumn{1}{c|}{1.35 } & 0.79 
\\ \cline{3-6} 
&   &FRCRN& \multicolumn{1}{c|}{6.20 } & \multicolumn{1}{c|}{1.47 } & 0.78 
\\ \cline{2-6} 
& \multirow{3}{*}{\cellcolor{gray!30}PnP-MVDR}  &\cellcolor{gray!30}DCCRNet
& \multicolumn{1}{c|}{\cellcolor{gray!30}6.68 } & \multicolumn{1}{c|}{\cellcolor{gray!30}1.48 } & \cellcolor{gray!30}\textbf{0.81} 
\\ \cline{3-6} 
&  \cellcolor{gray!30} &\cellcolor{gray!30}DCUNet
& \multicolumn{1}{c|}{\cellcolor{gray!30}\textbf{6.73 }} & \multicolumn{1}{c|}{\cellcolor{gray!30}\textbf{1.53} } & \cellcolor{gray!30}0.80 
\\ \cline{3-6} 
&  \cellcolor{gray!30}\multirow{-3}{*}{PnP-MVDR} &\cellcolor{gray!30}FRCRN& \multicolumn{1}{c|}{\cellcolor{gray!30}6.62 } & \multicolumn{1}{c|}{\cellcolor{gray!30}\textbf{1.53} } & \cellcolor{gray!30}0.80 
\\ \hline
\multirow{13}{*}{Factory} & Unprocessed  &-
& \multicolumn{1}{c|}{5.18} & \multicolumn{1}{c|}{1.11} &  0.56
\\ \cline{2-6} 
& MVDR  &-
& \multicolumn{1}{c|}{5.48 } & \multicolumn{1}{c|}{1.20 } & 0.69 
\\ \cline{2-6} 
& PIT  &-
& \multicolumn{1}{c|}{2.26} & \multicolumn{1}{c|}{1.16} &  0.53
\\ \cline{2-6} 
& Sepformer  &-
& \multicolumn{1}{c|}{6.18 } & \multicolumn{1}{c|}{1.24 } & 0.79 
\\ \cline{2-6} 
& DeepMVDR  &-
& \multicolumn{1}{c|}{2.53} & \multicolumn{1}{c|}{1.17} &  0.62
\\ \cline{2-6} 
& \multirow{3}{*}{Denoise+MVDR}  &DCCRNet
& \multicolumn{1}{c|}{4.47 } & \multicolumn{1}{c|}{1.17 } & 0.42 
\\ \cline{3-6} 
&    &DCUNet
& \multicolumn{1}{c|}{4.65 } & \multicolumn{1}{c|}{1.21 } & 0.45 
\\ \cline{3-6}
&    &FRCRN
& \multicolumn{1}{c|}{5.22 } & \multicolumn{1}{c|}{1.26 } & 0.44 
\\ \cline{2-6}
& \multirow{3}{*}{MVDR+Denoise}  &DCCRNet
& \multicolumn{1}{c|}{5.45 } & \multicolumn{1}{c|}{1.29 } & 0.81 
\\ \cline{3-6} 
&    &DCUNet
& \multicolumn{1}{c|}{5.49 } & \multicolumn{1}{c|}{1.37 } & 0.81 
\\ \cline{3-6} 
&    &FRCRN
& \multicolumn{1}{c|}{6.34 } & \multicolumn{1}{c|}{1.41 } & \textbf{0.82}
\\ \cline{2-6} 
& \multirow{3}{*}{\cellcolor{gray!30}PnP-MVDR}  &\cellcolor{gray!30}DCCRNet
& \multicolumn{1}{c|}{\cellcolor{gray!30}\textbf{7.23} } & \multicolumn{1}{c|}{\cellcolor{gray!30}1.50 } &  \cellcolor{gray!30}0.81 
\\ \cline{3-6} 
&   \cellcolor{gray!30} & \cellcolor{gray!30}DCUNet
& \multicolumn{1}{c|}{\cellcolor{gray!30}7.08 } & \multicolumn{1}{c|}{\cellcolor{gray!30}1.48 } &  \cellcolor{gray!30}0.80 
\\ \cline{3-6} 
&  \cellcolor{gray!30} \multirow{-3}{*}{PnP-MVDR}&\cellcolor{gray!30}FRCRN& \multicolumn{1}{c|}{\cellcolor{gray!30}7.11 } & \multicolumn{1}{c|}{\cellcolor{gray!30}\textbf{1.55} } &  \cellcolor{gray!30}\textbf{0.82}
\\ \hline
\multirow{13}{*}{Street} & Unprocessed  &
-
& \multicolumn{1}{c|}{5.65} & \multicolumn{1}{c|}{1.13} &  0.60
\\ \cline{2-6} 
& MVDR  &-
& \multicolumn{1}{c|}{5.57 } & \multicolumn{1}{c|}{1.25 } & 0.71 
\\ \cline{2-6} 
& PIT  &-
& \multicolumn{1}{c|}{3.15} & \multicolumn{1}{c|}{1.16} &  0.60
\\ \cline{2-6} 
& Sepformer  &-
& \multicolumn{1}{c|}{6.76 } & \multicolumn{1}{c|}{1.34 } & 0.79 
\\ \cline{2-6} 
& DeepMVDR  &
-& \multicolumn{1}{c|}{3.03} & \multicolumn{1}{c|}{1.17} &  0.63
\\ \cline{2-6} 
&\multirow{3}{*}{Denoise+MVDR}   &DCCRNet& \multicolumn{1}{c|}{4.05 } & \multicolumn{1}{c|}{1.15 } & 0.48 
\\ \cline{3-6} 
&    &DCUNet& \multicolumn{1}{c|}{4.53 } & \multicolumn{1}{c|}{1.24 } & 0.49 
\\ \cline{3-6}
&    &FRCRN& \multicolumn{1}{c|}{4.92 } & \multicolumn{1}{c|}{1.17 } & 0.47 
\\ \cline{2-6}
& \multirow{3}{*}{MVDR+Denoise}  &
DCCRNet
& \multicolumn{1}{c|}{5.08 } & \multicolumn{1}{c|}{1.25 } & 0.80 
\\ \cline{3-6} 
&    &
DCUNet
& \multicolumn{1}{c|}{5.29 } & \multicolumn{1}{c|}{1.39 } & \textbf{0.81} 
\\ \cline{3-6} 
&    &
FRCRN& \multicolumn{1}{c|}{6.34 } & \multicolumn{1}{c|}{1.37 } & 0.80 
\\ \cline{2-6} 
& \multirow{3}{*}{\cellcolor{gray!30}PnP-MVDR}  &
\cellcolor{gray!30}DCCRNet
& \multicolumn{1}{c|}{\cellcolor{gray!30}6.73 } & \multicolumn{1}{c|}{\cellcolor{gray!30}1.40 } & \cellcolor{gray!30}0.78 
\\ \cline{3-6} 
&  \cellcolor{gray!30}  &
\cellcolor{gray!30}DCUNet
& \multicolumn{1}{c|}{\cellcolor{gray!30}6.70 } & \multicolumn{1}{c|}{\cellcolor{gray!30}\textbf{1.47} } &  \cellcolor{gray!30}0.78 
\\ \cline{3-6} 
& \cellcolor{gray!30}  \multirow{-3}{*}{PnP-MVDR}&\cellcolor{gray!30}FRCRN& \multicolumn{1}{c|}{\cellcolor{gray!30}\textbf{6.87} } & \multicolumn{1}{c|}{\cellcolor{gray!30}1.46 } & \cellcolor{gray!30}0.80 
\\
\bottomrule
\end{tabular}
\end{table}

Hence, in order to explain these features of our framework, in this work, we apply 3 different kinds of DNN-based denoisers trained by different datasets. These denoisers are summarized in table~\ref{denoiser} and are introduced in detail as follows. 

\textit{1) DCUNet-Based Denoiser:} Deep Complex U-Net(DCUNet)\cite{dcunet} is an advanced neural network architecture specifically designed for speech enhancement in the time-frequency (T/F) domain. It extends the classic U-Net structure\cite{unet} to handle complex-valued spectrograms, enabling joint processing of both magnitude and phase information for high-quality audio reconstruction.In our work, we insert larger DCUnet-20 (Large-DCUnet-20)\footnote{The open-source DCUNet is available at https://huggingface.co/JorisCos/
DCUNet\_Libri1Mix\_enhsingle\_16k.} trained by libri1mix dataset into our framework since the authors in\cite{dcunet} claim that this network with more channels in each layer outperforms other architectures.

\textit{2) DCCRNet-Based Denoiser:} Deep Complex Convolution Recurrent Network (DCCRNet)~\cite{dccrnet} conducts complex valued speech enhancement via the convolutional encoder-decoder(CED) architecture with a complex LSTM layer between the encoder and the decoder. DCCRNet not only benefits from 2D convolution layer (Conv2d) to extract high-level features for better enhancement but also benefits from LSTM to model temporal context with less parameters. Here, we plug DCCRN-CL\footnote{The open-source DCCRNet is available at https://huggingface.co/JorisCos/
DCCRNet\_Libri1Mix\_enhsingle\_16k.} which is also trained by Libri1Mix dataset into our framework.

\textit{3) FRCRN-Based Denoiser:} Frequency Recurrent Convolutional Recurrent Networks (FRCRN)\cite{frcrn} is
trained to predict complex ideal ratio mask for target speech via an extended CED architecture, known as convolutional recurrent encoder-decoder structure which is able to improve feature representation. In addition, the recurrent module and attention block of FRCRN are designed to model the long-term temporal dependencies and to facilitate information flow respectively. The FRCRN model\footnote{The open-source FRCRN is available at https://www.modelscope.cn/mod\\els/damo/speech\_frcrn\_ans\_cirm\_16k.} we insert into our framework is trained by Deep Noise Suppression (DNS)-Challenge dataset\cite{dns}, of which 30\% of dataset contains reverberant components. 

Note that once the denoiser has been already pre-trained, it can be directly plugged into the proposed framework without additional modifications. 

In experiment, our objective is to enhance the performance of the vanilla MVDR beamforming in complex environments. Therefore, in our experiments, we consider the MVDR as the baseline for comparison. Also, since our experimental scenarios involve additive noise, we also explore MVDR+Denoise and Denoise+MVDR, which will be explained in detail next, to ensure a fair comparison. After everything is set up, we present all the comparison methods as follows.

\begin{itemize}
\item \textbf{MVDR:} The spatial selectivity of MVDR is used to enhance the speech source signals in specific directions and suppress the signals in other directions for the purpose of extraction and separation. We also consider the MVDR as the baseline for comparison.
\item \textbf{PIT:} PIT~\cite{pit} is a single-channel speech separation network for solving label alignment problems by minimizing separation errors. Here, we directly use the pretrained model\footnote{The open-source PIT model is available at https://github.com/JusperLee/U
tterancePIT-Speech-Separation.} to inference.
\item \textbf{Sepformer:} Sepformer\cite{sepformer} is a Transformer-based DNN for speech separation that uses dual-path self-attention to efficiently separate overlapping speakers. It outperforms traditional methods by learning long-range dependencies and generating high-quality masks in an end-to-end manner. We directly use the pretrained model\footnote{The open-source Sepformer model is available at https://huggingface.co/sp
eechbrain/sepformer-whamr16k.} to inference.
\item \textbf{DeepMVDR:} The interference covariance matrix is obtained employing a mask-based DNN in the time-frequency domain(we used the PIT here), and subsequently employed to implement the MVDR beamforming.
\item \textbf{Denoise+MVDR:} This is the first intuitive connection strategy that combines physics-based and data-driven approaches. To be specific, the multi-channel speech signal is first enhanced channel-by-channel using the denoisers, and then the MVDR beamforming is applied.
\item \textbf{MVDR+Denoise:} The other intuitive connection strategy to integrate physics-based and data-driven approaches, specifically, MVDR beamforming is applied first and then a denoiser is used to perform post-filtering. Otherwise, we take it as the second baseline to compare the effects of different combination strategies.
\item \textbf{PnP-MVDR:} Our proposed PnP strategy for speech separation with different denoisers.
\end{itemize}

\subsection{Evaluation Metrics}
For evaluation metrics, we used the frequency-weighted segmental signal-to-noise ratio (fwSNRseg)~\cite{fwSNRseg}, the perceptual evaluation of speech quality (PESQ)~\cite{pesq} and the short-time objective intelligibility (STOI)~\cite{stoi} to evaluate the performance of speech separation at the signal level. PESQ is a speech quality metric that takes phase information into account. STOI assesses the intelligibility of separated speech by measuring the correlation between the temporal envelopes of the separated signal and its clean reference. fwSegSNR quantifies the signal difference between separated and clean speech in a reweighted frequency domain, where the frequency bin weights are determined by the amplitude of the clean signal.

\section{Experimental Results}

\begin{figure*}[t!]
   \includegraphics[width=1.0\textwidth, trim = 1.2cm 3cm 0cm 5cm, clip]{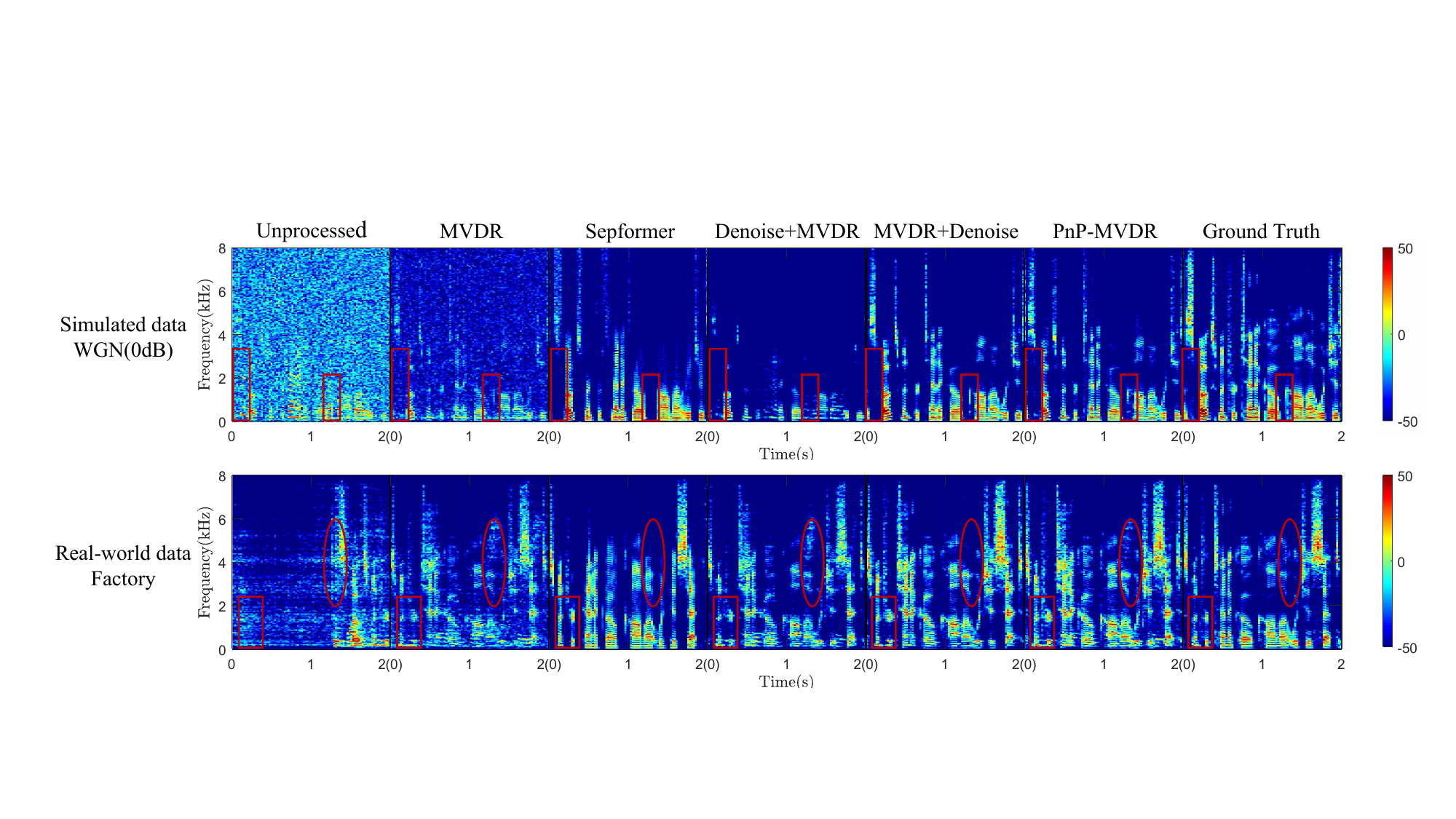}
\caption{{The visualization results of all comparison methods with the speech truncated to 2 s. From left to right, we present the spectrograms of unprocessed speech, and the estimated speech processed by the vanilla MVDR, Sepformer, MVDR+Denoise, Denoise+MVDR, PnP-MVDR and ground truth respectively.}}
\label{fig:spectrogram}
\end{figure*}

Table~\ref{simulated} and table~\ref{real-world} summarize
the comparison results of all evaluation metrics on simulated data and real-world data respectively. From both tables, we can directly see that the proposed method stands out prominently in all experimental scenarios.  For visual comparison, we also take the simulation scenarios SNR = 0 dB of WGN and real-world scenarios of factory as examples, and draw the speech spectrograms of the results of all the methods, depicted in Fig.~\ref{fig:spectrogram}. From the figure, the spectrogram of PnP-MVDR preserves the essential features of speech more completely, which verifies the rationality of our motivation and modeling process, and indicates the robustness in the strong interference scenario. 
This is evidenced by the significant variations exhibited in the region delineated by the red solid line. In the following, we will compare the performance of different methods from various aspects to further demonstrate the superiority of our  proposed framework.    

\begin{figure*}[t!]
\centering
   \includegraphics[width=1\textwidth, trim=4cm 7cm 4cm 6cm, clip]{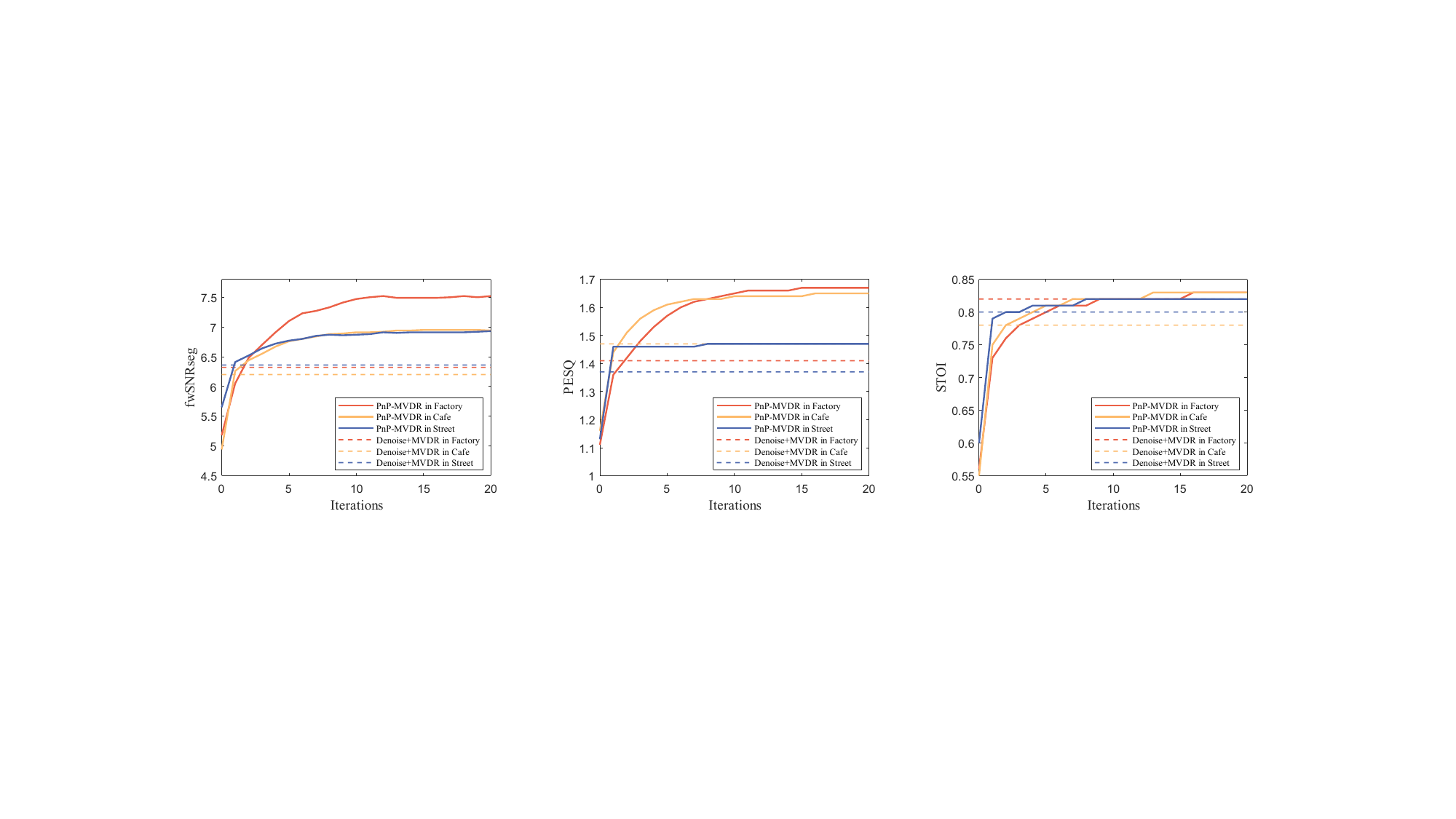}
\caption{{The results for PnP-MVDR methods across iterations. The experimental scenario is a real scene containing three types of noise from factory cafe street and the denoiser used is FRCRN.}}
\label{fig:iters}
\end{figure*}

\noindent\textbf{Comparison of MVDR, MVDR+Denoise and Denoise+MVDR:} Although MVDR effectively suppresses noise and interference signals by minimizing output power, its optimization process does not account for the inherent structural characteristics of speech signals. As illustrated in section \uppercase\expandafter{\romannumeral 3}, this limitation arises from the lack of explicit speech signal modeling in the MVDR framework. Compared with MVDR, the MVDR+Denoise method introduces an additional denoising stage to process the MVDR output. Theoretically, this post-processing step can enhance overall performance by suppressing residual spectral distortions that are treated as noise components in the MVDR output signal. 

As shown in Tables~\ref{simulated} and \ref{real-world}, although there is no significant difference in the fwSNRseg scores between MVDR and MVDR+Denoise, the STOI and PESQ scores do. This is attributed to the fact that the denoise operation repairs the speech spectrum and is more conducive to improving speech intelligibility.

In contrast to MVDR+Denoise, Denoise+MVDR first applies the DNN-based denoiser to multi-channel input signals. The core issue is that deep neural networks are fundamentally designed to approximate complex mappings between input data and target labels, which inherently leads to a phenomenon that phase information of the observed signal is corrupted. However, since MVDR fundamentally depends on spatial correlation, the degradation of phase information can significantly compromise its performance. The experimental results aligned with our expectations, demonstrating that Denoise+MVDR consistently underperformed MVDR+Denoise across all evaluation metrics. 
\begin{figure}[t!]
\centering
   \includegraphics[width=0.5\textwidth, trim = 7cm 2.5cm 7cm 2.5cm, clip]{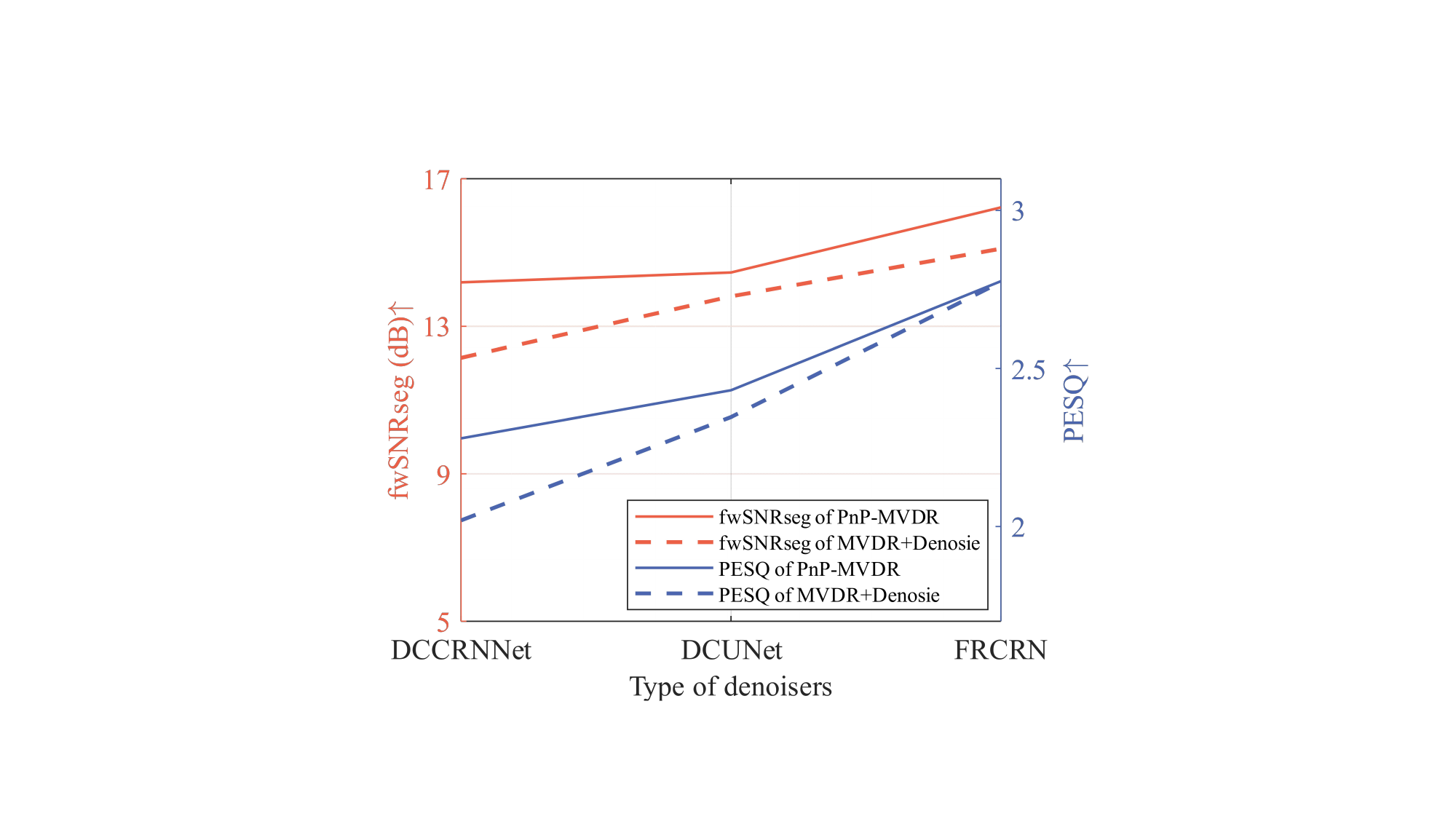}
\caption{{The fwSNRseg and PESQ results of MVDR+Denoise and PnP-MVDR by applying different denoisers on simulated data. The metrics are the results after averaging all the scenes.}}
\label{fig:denoiser}
\end{figure}

\noindent\textbf{Comparison of MVDR+Denoise and PnP-MVDR:} 
The two solutions both combine model-based and data-driven approaches, differentiated by their structural organization: a cascaded implementation in the first case, contrasted with our PnP architecture in the proposed method. Our framework employs data-driven methods to extract speech structural features that constrain output signals to maintain speech intrinsic properties, with iterative enhancement through gradual incorporation of speech priors. Fig.~\ref{fig:iters} presents  the improved performance of PnP-MVDR with an increasing number of iterations across all evaluated metrics. 


In addition, our experiments employed three denoisers with varying properties to implemented PnP-MVDR. As shown in the table~\ref{simulated} and~\ref{real-world}, these denoisers exhibit a measurable influence on separation performance. To better demonstrate the impact, we use the output of MVDR+Denoise as a benchmark for evaluating denoisers' performance, as it applies denoising operation to the same MVDR results. This allows for clear observation of how different denoisers affect our proposed PnP-MVDR. Based on the above analysis, we draw a line graph of the correlation between PnP-MVDR and MVDR+Denoise on as shown in Fig.~\ref{fig:denoiser}. We find that as the performance of Denoise + MVDR improves, PnP-MVDR simultaneously demonstrates a gradual upward trend. On real-world data, although fwSNRseg does not show a pronounced trend, the performance tendency is still reflected in the PESQ scores. All of these imply that the denoiser has a positive effect on PnP-MVDR, reflecting the flexibility of our framework. This is because the performance of our proposed method is based on the ability of the denoiser to extract a speech priors, i.e., our framework is flexible, as well as updatable with advances of denoiser.

\begin{figure}[t!]
\centering
\includegraphics[width=0.5\textwidth, trim = 10cm 3cm 10cm 2.5cm, clip]{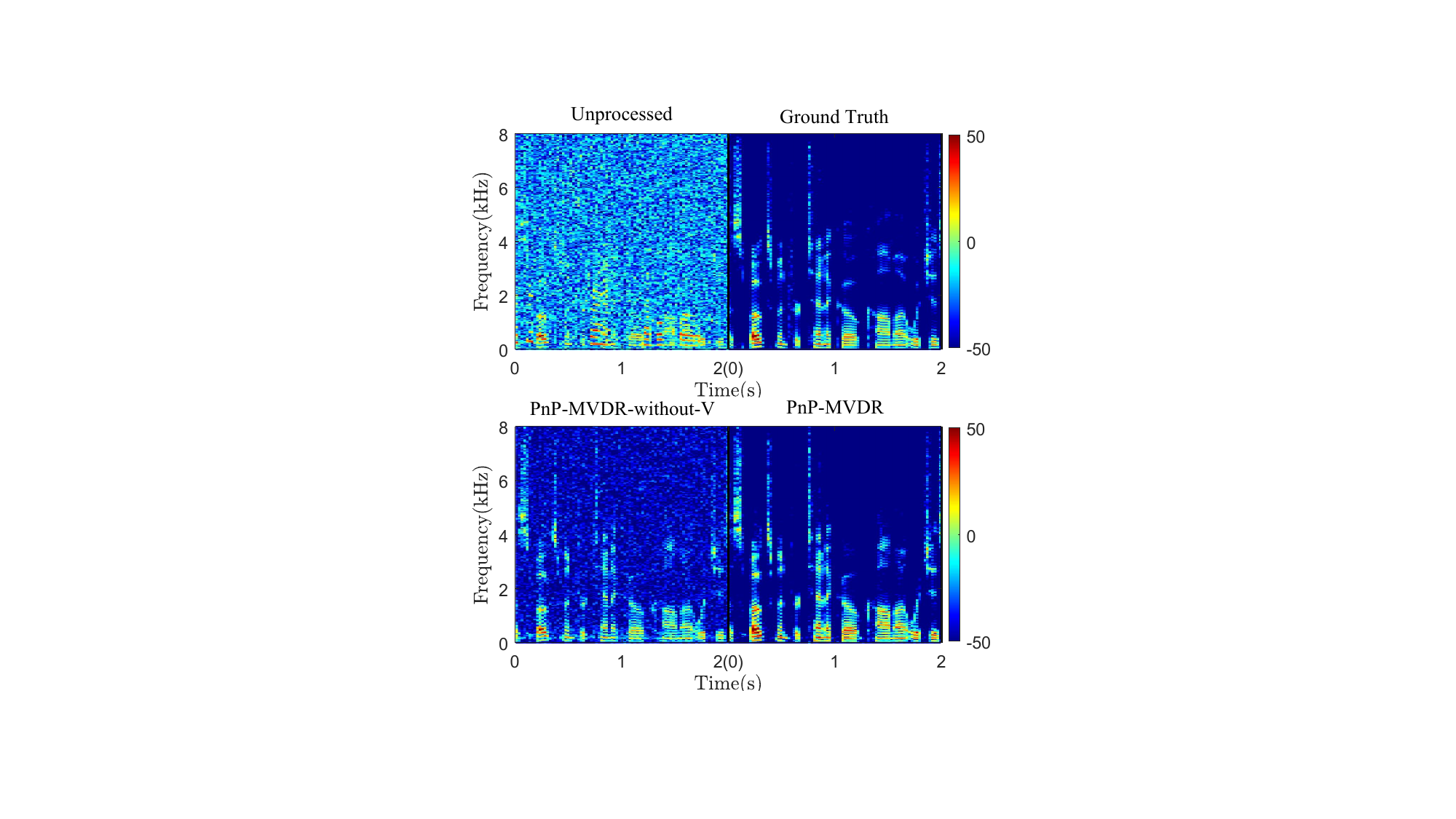}
\caption{{The visualization results of PnP-MVDR and PnP-MVDR-without-V under
the scenario with WGN at SNR = 0 dB}}
\label{fig:without_V}
\end{figure}

\noindent\textbf{Comparison of PnP-MVDR and Sepformer:}
We also compared our PnP framework with an end-to-end approach, which learns the mapping of mixed signals to separated signals by training large amounts of data. However, it exhibits limited generalization capability and sensitivity to noise diversity. As evidenced in Table \ref{simulated} and~\ref{real-world}, while Sepformer demonstrates competitive performance in little cases, it generally underperforms PnP-MVDR and shows inconsistent results across scenarios.This further validates the importance of our modeling based on the physical method.

\noindent\textbf{Discussion of $V(n,k)$:}
To explore the capability of term $V(n,k)$, we perform a comparison experiment between PnP-MVDR and PnP-MVDR without term $V(n,k)$ (termed PnP-MVDR-without-$V$), where the denoiser used here is FRCRN. In theory, without explicit noise and interference modeling, PnP-MVDR-without-$V$ is unable to remove the additive noise from the observed speech. Fig.~\ref{fig:without_V} present the spetrogram of  PnP-MVDR and PnP-MVDR-without-$V$.  From Fig.~\ref{fig:without_V}, it is clear that PnP-MVDR-without-V fails to remove the additive noise from the observed speech. This is strong validation that introducing the auxiliary term $V(n,k)$ to our framework is reasonable. Further, from this comparison experiments, we can deduce that in addition to incorporating sophisticated speech prior for the framework, the capability of the DNN-based denoiser module in our proposed method is to eliminate some unknown and un-modelled noises, such as structural noise during the optimization processing, instead of additive background noises~\cite{zhang2021plug}.

\section{Conclusion}

In this paper, we presented a Plug-and-Play MVDR (PnP-MVDR) beamforming framework that integrates data-driven speech priors into the conventional MVDR structure via a regularization-by-denoising strategy. The proposed approach enhanced robustness in complex acoustic environments while preserving interpretability. Furthermore, a complete microphone beamforming system with a custom linear array and cooperative DSP-NPU architecture was developed, validating the method's practical feasibility. Evaluations on both simulated and real-world data demonstrated that PnP-MVDR consistently outperforms existing methods, offering an effective solution for hardware-friendly speech signal processing.

\bibliographystyle{IEEEtran}
\bibliography{reference}

\newpage


\vfill

\end{document}